\documentclass[epj]{svjour}
\usepackage[british]{babel}
\usepackage{epsfig}
\usepackage{appendix}
\usepackage{subfig}
\usepackage[T1]{fontenc}

\usepackage[switch,columnwise]{lineno}

\usepackage[space]{cite}
\usepackage{amsmath}
\usepackage{multirow}

\newcommand\ba{\begin{eqnarray}}
\newcommand\ea{\end{eqnarray}}

\newcommand{\be}{\begin{equation}}
\newcommand{\ee}{\end{equation}}

\newcommand\bi{\begin{itemize}}
\newcommand\ei{\end{itemize}}
\newcommand\bc{\begin{center}}
\newcommand\ec{\end{center}}

\usepackage{xspace}
\begin{document}
\title{Measurement of neutron and proton analyzing powers on $C$, $CH$, $CH_2$ and $Cu$ targets in the momentum region 3-4.2 GeV/c}

\author{
%The ALPMO2 Collaboration\\ 
S.N. Basilev\inst{1} 
\and
 Yu.P. Bushuev\inst{1} 
 \and
 O.P. Gavrishchuk\inst{1} 
 \and
  V.V. Glagolev\inst{1} 
\and
 D.A. Kirillov\inst{1} 
 \and
 N.V. Kostayeva\inst{1}  
 \and
 A.D. Kovalenko\inst{1}  
 \and
 K.S. Legostaeva\inst{1} 
 \and
 A.N. Livanov\inst{1} 
 \and
I.A. Philippov\inst{1} 
\and
 N.M. Piskunov\inst{1} 
 \and
 A.A. Povtoreiko\inst{1} 
 \and
 P.A. Rukoyatkin\inst{1} 
\and
 R.A. Shindin\inst{1} 
 \and
 A.V. Shipunov\inst{1} 
 \and
 A.V. Shutov\inst{1} 
 \and
 I.M. Sitnik\inst{1} 
 \and
 V.M. Slepnev\inst{1} 
 \and
 I.V. Slepnev\inst{1} 
 \and
 A.V. Terletskiy\inst{1} 
\and 
K. Hamilton\inst{2}
\and 
 R. Montgomery\inst{2}
 \and 
J. R.M. Annand\inst{2}
\and 
D. Marchand\inst{3}
\and 
 Y. Wang\inst{3}
\and 
E. Tomasi-Gustafsson\inst{4}
\and
C.F. Perdrisat\inst{5}
 \and 
 V. Punjabi\inst{6} 
\and 
G. Martinska\inst{7}
\and J. Urban\inst{7}
\and 
J. Mu\v{s}insky\inst{8}
% etc
}                     % Do not remove
\institute{Joint Institute for Nuclear Research, 141980 Dubna, Russia
\and 
University of Glasgow, Glasgow G12 8QQ, Scotland, UK
\and 
IPN Orsay, Universit\'e Paris-Saclay, 91406 ORSAY, France 
\and 
IRFU, CEA, Universit\'e Paris-Saclay, 91191 Gif-sur-Yvette, France
\and 
The College of William and Mary, Williamsburg, VA 23187, USA
\and 
Norfolk State University, Norfolk, VA 23504, USA
\and 
University of P.J. Safarik, Jesenna\ 5, SK-04154 Ko\v{s}ice, Slovak Republic
\and 
Institute of Experimental Physics, Watsonova 47, SK-04001 Ko\v{s}ice, Slovak Republic
}
\date{Received: date / Revised version: date}
\abstract{
The analyzing powers for proton elastic scattering ($\vec p A\to pX$) and neutron charge exchange ($\vec n A\to p X$) reactions on nuclei have been measured on $ C$, $CH$, $CH_2$ and $Cu$ targets at incident neutron momenta 3.0 - 4.2 GeV/c by detecting one charged particle in forward direction. The polarized neutron measurements are the first of their kind. The experiment was performed using the Nuclotron accelerator in JINR Dubna, where polarized neutrons and protons were obtained from breakup of a polarized deuteron beam which has a maximum momentum of 13 GeV/c. The polarimeter ALPOM2 was used to obtain the analyzing power dependence on the transverse momentum of the final-state nucleon. These data have been used to estimate the figure of merit of a proposed experiment at Jefferson Laboratory to measure the recoiling neutron polarization in the quasi-elastic $^2H(\vec e,e'\vec n)$ reaction, which yields information on the charge and magnetic elastic form factors of the neutron.
\PACS{
      {PACS-key}{discribing text of that key}   \and
      {PACS-key}{discribing text of that key}
     } % end of PACS codes
} %end of abstract
\titlerunning{Neutron and proton analyzing powers on $C$, $CH$, $CH_2$ and $Cu$ in the GeV region}
\maketitle
\section{Introduction}
\label{intro}

Polarimetry of nucleons in the GeV region requires the measurement of the azimuthal distribution resulting from a secondary scattering of the polarized nucleon in a suitable  analyzing material. $C$ or $CH_2$ are often used as the analyzers in proton polarimeters. Analyzing power measurements were made at Saclay \cite{Cheung:1995ei} (and references therein)  and Dubna \cite{Azhgirey:2004yk}, using thick analyzers, and as part of a program of study of elastic and quasi-elastic  $\vec d p$ reactions \cite{Azhgirey:2008} up to incident momenta of several GeV/c. New measurements at Dubna will extend the incident momenta up to ~7.5 GeV/c \cite{ALPOM2}.

%%For protons, accurate data for the analyzing power on CH2 and CH targets are necessary for these experiments, requiring the measurement of the polarization of protons and neutrons in nuclear reactions. Such data have been obtained in Saclay \cite{Cheung:1995ei} (and references therein) and Dubna \cite{Azhgirey:2004yk}, using thick analyzers, and as part of a program of study of elastic and quasi-elastic  $\vec d p$ reactions \cite{Azhgirey:2008}. A measurement of the angular distribution of the analyzing power of CH2 for protons to as high a momentum as possible, at least up to 7.5 GeV/c is of the greatest interest.

Up to now neutron polarimetry has generally been based on free elastic $np$ scattering
or elastic-like $np$ scattering from nuclei. The kinematic reconstruction of the scattering process is highly desirable to select the range of polar scattering angles where the analyzing power is relatively large. This may be achieved by using an active, position-sensitive analyzer to detect the recoiling proton and thus localise the interaction position of the incident neutron. Alternatively charge-exchange neutron scattering may be used, where the trajectory of the energetic, forward-angle proton can be tracked. 

In comparison to proton scattering, the analyzing power $A_{y}$ for
polarized neutron scattering at GeV energies is poorly known. Analyzing powers for $n+p\to n+p$ and $n+p\to p+n$ scattering exist only for thin $^1H$ or $^2H$ targets.
%%Cross section and analyzing powers for $np$, both for elastic and charge exchange reactions, are known up to 29 GeV/c. No data are known to exist for thick analyzers, made of scintillator material. 
Free $np$ scattering is in principle the best analyzer of neutron polarization,
but the use of a hydrogen analyzer is challenging technically and
up to now scattering from $C$, $CH$ or $CH_2$ 
has most commonly been used. However $A_{y}$ for elastic-like scattering
from nuclei is unknown in the few-GeV energy range and will be lower than for the free-scattering case. 

% from John intro
Polarized deuteron beams up to 13 GeV/c momentum, producing polarized neutron and protons after break-up, are presently available only  at the Nuclotron accelerator in Dubna. An important upgrade of the accelerator complex has been recently undertaken \cite{Kekelidze:2017ghu}, as first step of the future NICA collider \cite{Kekelidze:2017ual}.
This resulted in a new program by the ALPOM2 collaboration, with the aim to measure analyzing powers up to the largest available momenta  \cite{ALPOM2}. The experimental program and the first results with polarized neutrons are reported here. 

These data are particularly important for pursuing the nucleon electromagnetic form factor measurements at JLab.

%%%%%%%%%%%%%%%%%%%%%%%%%%%%%
\subsection{Proton form factor measurements}
%%%%%%%%%%%%%%%%%%%%%%%
The electromagnetic form factors (EMFF) of elastic electron-Nucleon ($eN$) scattering are representative of the charge and the magnetic currents of the nucleons (for recent reviews on electromagnetic form factors see Refs. \cite{Pacetti:2015iqa,Punjabi:2015bba}).  
 
Early determinations of the EMFF used Rosenbluth separation of differential cross section measurements  \cite{Pacetti:2015iqa}, but as electron accelerator technology has developed to produce high-current, high polarization, continuous-wave electron beams, the measurement of polarized observables has become the method of choice. Polarization observables are especially useful in separating a small amplitude from an otherwise dominant one and are believed to be relatively insensitive to radiative correction effects. 

The use of polarization was first proposed by the Kharkov school \cite{Akhiezer:1968ek,Akhiezer:1974em} as an alternate method to Rosenbluth separation to determine the nucleon EMFF  from  polarized elastic  electron-proton ($\vec e p$) and electron-neutron ($ \vec e n$) scattering (where the neutrons are bound in  a deuteron or $^3\!He$ target). The ratio of the longitudinal to transverse recoil nucleon polarization in elastic $eN$ scattering, with longitudinally polarized electrons, is proportional to the ratio of electric to magnetic form factors, whereas the unpolarized cross section  depends on the squares of form factors. These types of double-polarization experiments,  denoted, in the following, {\it recoil polarization experiments }, require the measurement of the polarization of the recoiling nucleon in elastic $eN$ scattering. Experiments of this type started in the 1990's (see below) when high-intensity, highly-polarized, high-duty-cycle electron beams became available.

Subsequently, the range of the squared four-momentum transfer, $Q^2$, has been extended at Jefferson Lab (JLab), using a polarized electron beam of energy up to 5.7 GeV to measure the four form factors of elastic eN scattering (electric and magnetic, for proton and neutron, $G^p_E$, $G^p_M$, $G^n_E$, $G^n_M$). Proton data have produced unexpected and intriguing results.  Contrary to the generally accepted scaling relation that the electric to magnetic form factor ratio $\mu_p G^p_E/G^p_M \sim 1$ ($\mu_p=2.79\!\mu_N$ is the proton anomalous magnetic moment) the recoil polarization data show an approximately linear decrease of this ratio,  clearly indicating that the electric and magnetic form factor have different dependence on $Q^2$, and therefore that the radial 
distributions of charge- and magnetization, are not the same.  This was an unexpected 
result and the various papers publishing these results 
\cite{Jones:1999rz,Punjabi:2005wq,Gayou:2001qd,Puckett:2011xg,Puckett:2010ac,Puckett:2017flj,Meziane:2010xc} have been cited more than 2000 times. 

JLab has recently undergone an energy upgrade, and is producing polarized beams of energy up to 11 GeV. Extending the $Q^2$ range of  $\mu_p G^p_{E}/G^p_{M}$ measurements should determine if the ratio does cross zero, as suggested by the existing data, while for $\mu_n G^n_{E}/G^n_{M}$ a completely virgin territory will be explored. With an increase in $Q^2$  the momentum of the recoiling nucleon also increases. Approval of the most recently published $G^p_E/G^p_M$ experiment \cite{Puckett:2017flj} relied heavily on measurements previously made at Dubna \cite{Azhgirey:2004yk}, which showed that the proton analyzing power at high momentum is sufficiently high to obtain a good precision for the EMFF ratio. Therefore the extension of the analyzing power data base to higher nucleon momenta is urgently needed, both for protons and, more critically, for neutrons where multi-GeV/c data is extremely limited. The status and the planned experiments for nucleon EMFF
are briefly reviewed here.

\begin{figure}
\centering
%\begin{sidewaysfigure}
%\resizebox{0.3\textwidth}{!}{
% \includegraphics{gepgmp_allpol_final_lin_lin.pdf}
 %}
 \includegraphics[width=7cm,angle=90]{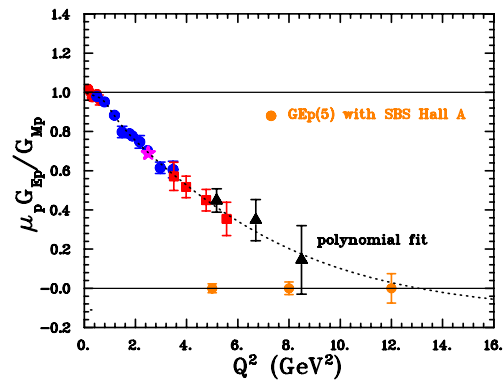}
\caption{Projections of experimental uncertainties for future JLab experiment E12-07-109 \protect\cite{PR12-07-109} (filled orange circles). 
Also shown for comparison are previous JLab $G^p_E/G^p_M$ data for 
GEp(1), GEp(3), and GEp$2\gamma$ \cite{Jones:1999rz,Punjabi:2005wq,Gayou:2001qd,
Puckett:2011xg,Puckett:2010ac,Puckett:2017flj,Meziane:2010xc}. }
\label{fig:Gep} 
%\subfigure
 %\end{sidewaysfigure}
\end{figure}

Experiment E12-07-109 \cite{PR12-07-109} will measure the proton form factor ratio up to $Q^2 = 12$~  (GeV/c)$^2$, and a projection of the achievable precision is displayed in Fig. \ref{fig:Gep}. To achieve this,  a high luminosity of $10^{39}$ cm$^{-2}$s$^{-1}$ and a large acceptance detector are necessary. The electron arm will consist in a Pb-Glass calorimeter, while the proton recoil polarimeter will be part of the  Super Bigbite Spectrometer (SBS) spectrometer. The SBS is equipped with a large-aperture dipole magnet of ~1.7 Tm integrated field strength, two 60 cm $CH_2$ analyzer blocks, ten planes of GEM chambers for proton tracking before and after the analyzers and a hadron calorimeter to select energetic protons in the experimental trigger. 

\subsection{Neutron Form factor measurements}
\begin{figure}
\centering
\includegraphics[width=7.cm,angle=90]{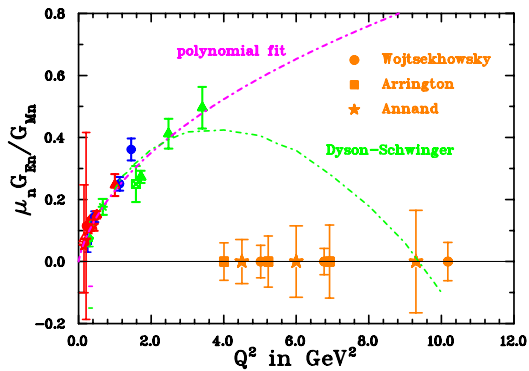}
\caption{The future data points proposed for recoil polarization $G^n_{E}/G_M^n$ experiments at JLab \protect\cite{PR12-11-009,PR12-17-004},  
and the JLab polarized $^3\!He$ experiment  \protect\cite{PR12-09-016}. 
Also shown are the 
data from Becker (filled green diamonds) \protect\cite{Becker:1999tw},
Eden (filled blue squares) \protect\cite{Eden:1994ji},
Glazier (red diamonds with internal +) \protect\cite{Glazier:2004ny},
Golak (empty green diamonds) \protect\cite{Golak:2000nt},
Herberg (empty red triangles) \protect\cite{Herberg:1999ud},
Meyerhof (filled green stars) \protect\cite{Meyerhoff:1994ev},
Ostrick (filled red diamonds) \protect\cite{Ostrick:1999xa}, and
Passchier (filled red circles) \protect\cite{Passchier:1999ju}. }
\label{fig:Gen} 
\end{figure}
Measurements of neutron elastic form factors generally use quasi-elastic 
$(e,e'n)$ scattering from the neutron embedded in a $^2\!H$
or $^3\!He $ target, where the scattered electron and recoiling
neutron are detected in coincidence. Quasi-elastic 
 scattering
is separated from inelastic processes on the basis of the measured
electron momentum and the angular correlation of the $e'$ and $n$.

The first recoil polarization  measurement of $G_{E}^{n}/G_{M}^{n}$ was made
at Bates Laboratory \cite{Eden:1994ji}, but the achieved precision was
limited, mainly due to lack of a Continuous Wave (CW) electron beam.
Following on from this, a polarized CW 855~MeV electron beam became
available at the MAMI accelerator in Mainz and a series of recoil polarization  measurements
were made with an unpolarized, liquid $D_2$ target \cite{Herberg:1999ud,Ostrick:1999xa},
in parallel with measurements using a polarized $^3\!He$ target. 
Subsequent measurements at MAMI were made in the A1 spectrometer hall
\cite{Glazier:2004ny} using a high resolution magnetic spectrometer to detect
the $e'$ and a more compact plastic-scintillator, neutron polarimeter.
At MAMI, recoil polarization  measurements extended to ${Q}^{2}=0.8$~(GeV/c)$^{2}$,
where ${Q}^{2}$ is limited by the maximum electron beam energy (currently
1.6 GeV). In JLab Hall-C, using the 6 GeV CEBAF accelerator, recoil polarization  measurements extended up to $Q^2$ = 1.45 (GeV/c)$^2$ \cite{Plaster:2005cx,Madey:2003av}. Using a polarized $^3\! He$ target in Hall-A \cite{Riordan:2010id}, a maximum value of $Q^2$ = 3.4 (GeV/c)$^2$ was achieved.

Two new recoil polarization  measurements of $G_{E}^{n}/G_{M}^{n}$ have been
proposed to run at the upgraded 11~GeV CEBAF accelerator. The first  
follows on from the previous Hall-C experiments \cite{PR12-11-009}, while the second in Hall-A \cite{PR12-17-004} proposes a polarimeter to measure charge-exchange
neutron scattering from Cu. The present charge exchange results, used to estimate (see below) the figure of merit of the polarimeter and hence the obtainable experimental precision, played a crucial role in the approval of this new polarimetry technique for E12-17-004.
%%%%%%%%%%%%%%%%%%%%%%%%%%%%%%%%%%%%%%%%%%%%%%%%%%%%%%%%%%
%%%%%%%%%%%%%%%%%%%%%%%%%%%
\subsection{Polarimetry in the GeV region}
\begin{figure}
\begin{centering}
\includegraphics[width=0.9\columnwidth]{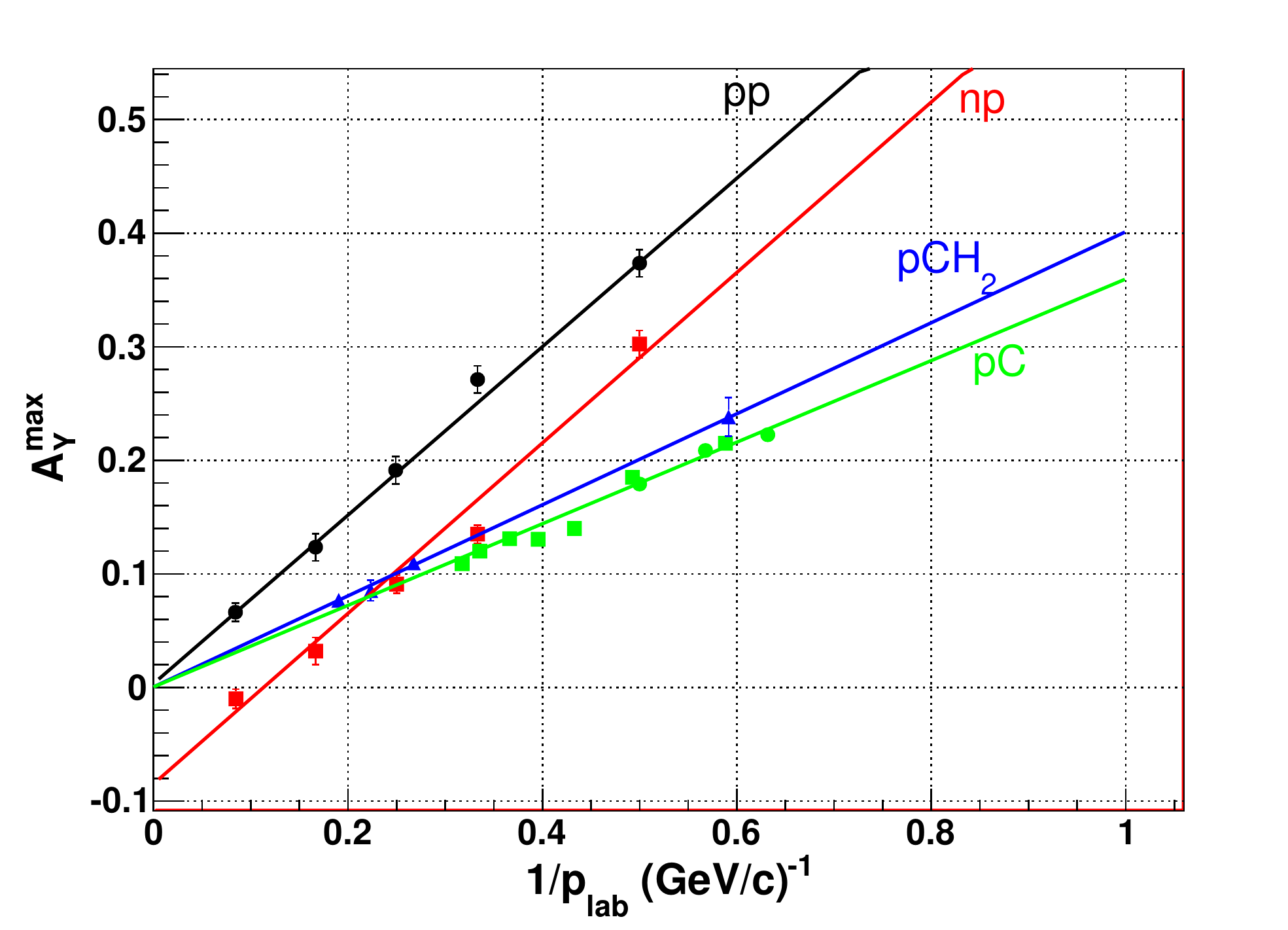} 
\par\end{centering}
\caption{\label{fig:Dep-Ay1}The dependence of the maximum of $A_{Y}$ on the inverse of $p_{lab}$, the momentum of the proton entering the polarimeter in the laboratory system. 
Black circles: ANL $d(\vec{p},p')n$ data \cite{PhysRevLett.35.632,Kramer:1977pf};
black line: linear fit. Red squares: ANL $d(\vec{p},n)p$ data \cite{Cheung:1995ei,Alekseev:1999ag} ;
red line: linear fit. Blue triangles \cite{Azhgirey:2004yk}: $\vec{p}+ CH_{2} \to charged\,+\,X$;
blue line: linear fit \cite{Azhgirey:2004yk}. Green squares \cite{Cheung:1995ei}
and circles \cite{Alekseev:1999ag}: $\vec{p}+ C\to charged\,+\,X$;
green line: linear fit \cite{Azhgirey:2004yk}. }
\end{figure}

%%%%%%%%%%%%%%%%%%%%%%%%%
The information on polarized nucleon scattering for incident momenta $p_{lab}\ge 1.5$~GeV/c (Fig. \ref{fig:Dep-Ay1})
comes from a number of sources. Measurements of the asymmetries of the $d(\vec{p},p')n$ and $d(\vec{p},n)p$
processes have been performed in the 1970s up to an incident momentum of 11~GeV/c \cite{PhysRevLett.35.632,Kramer:1977pf}. The  $d(\vec{p},p')n$ data are consistent with elastic $\vec{p}+ p \to  p+p$
measurements \cite{Spinka:1983rz} so that these experiments are equivalent to free polarized $pp$
and $pn$ scattering. Inclusive measurements of $\vec{p}+CH_{2} \to p+X$~\cite{Azhgirey:2004yk},
and $\vec{p}+ C \to p+X$ have
been obtained for the calibration of proton polarimeters used at ANL,
JINR Dubna and JLab ~\cite{Cheung:1995ei,Alekseev:1999ag}.

\begin{figure}[htb]
\begin{centering}
\includegraphics[width=1\columnwidth]{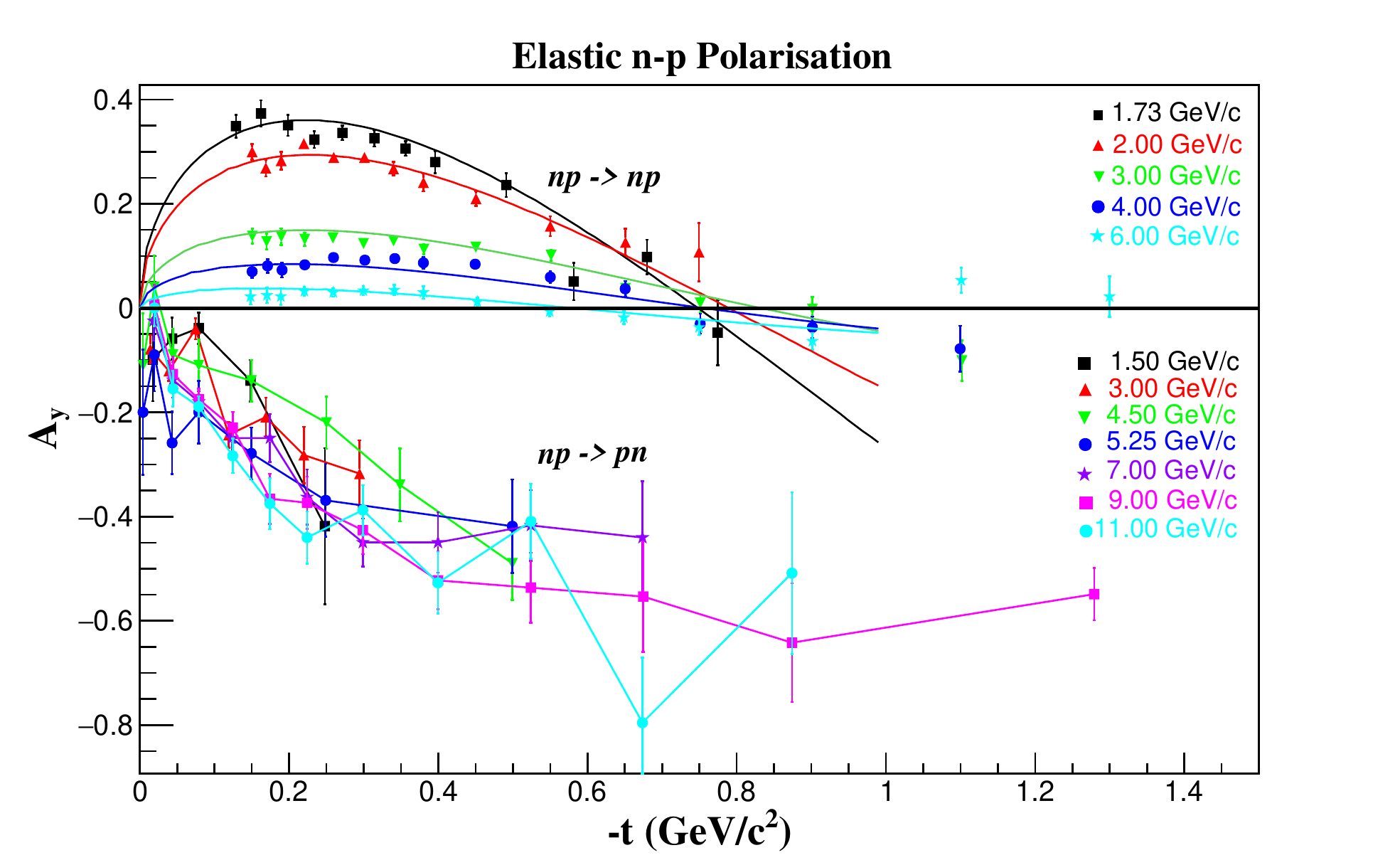}
\par\end{centering}
\caption{\label{fig:Dep-Ay}Top:  $t$-dependence of the
polarisation of $np$ scattering  for different values of $p_{lab}$ \cite{PhysRevLett.35.632,Kramer:1977pf}. The smooth
dotted lines show the fit of Ref. \cite{Ladygin:409018} to the $np$ data.
Bottom: the $t$ dependence of charge-exchange $np$
scattering for different values of $p_{lab}$ \cite{PhysRevLett.30.1183,Robrish:1970jw}. The color coding relates the data
to momentum labels. }
\end{figure}

The data displayed in Fig. \ref{fig:Dep-Ay1} show an approximately 
linear dependence of the maximum value of the angle dependent analyzing power, $A_{y}^{max}$, on $1/p_{lab}$, but there is
a significant negative offset of the $pn$ data with respect to $pp$. There is a factor of two reduction in the analyzing power of $\vec{p}+^{12}\!C$ with respect to free $pp$ scattering, but to our knowledge there
are no data on polarized $np$ scattering from nuclei in the multi-GeV
energy domain.

Measurements of the asymmetries of polarized charge exchange $n+\vec{p}\to p+X$
scattering, displayed in Fig. \ref{fig:Dep-Ay} up to 11~GeV/c,  have also been made at ANL in
the 1970s \cite{Robrish:1970jw,PhysRevLett.30.1183}.
While the $np$ (equivalent to $pn$) polarization is strongly dependent on the incident nucleon momentum $p_{lab}$, charge-exchange $np$ displays no apparent (given the spread in the data) strong dependence of $A_{y}$ on $p_{lab}$.

\begin{figure}
\begin{center}
\includegraphics[width=0.95\columnwidth]{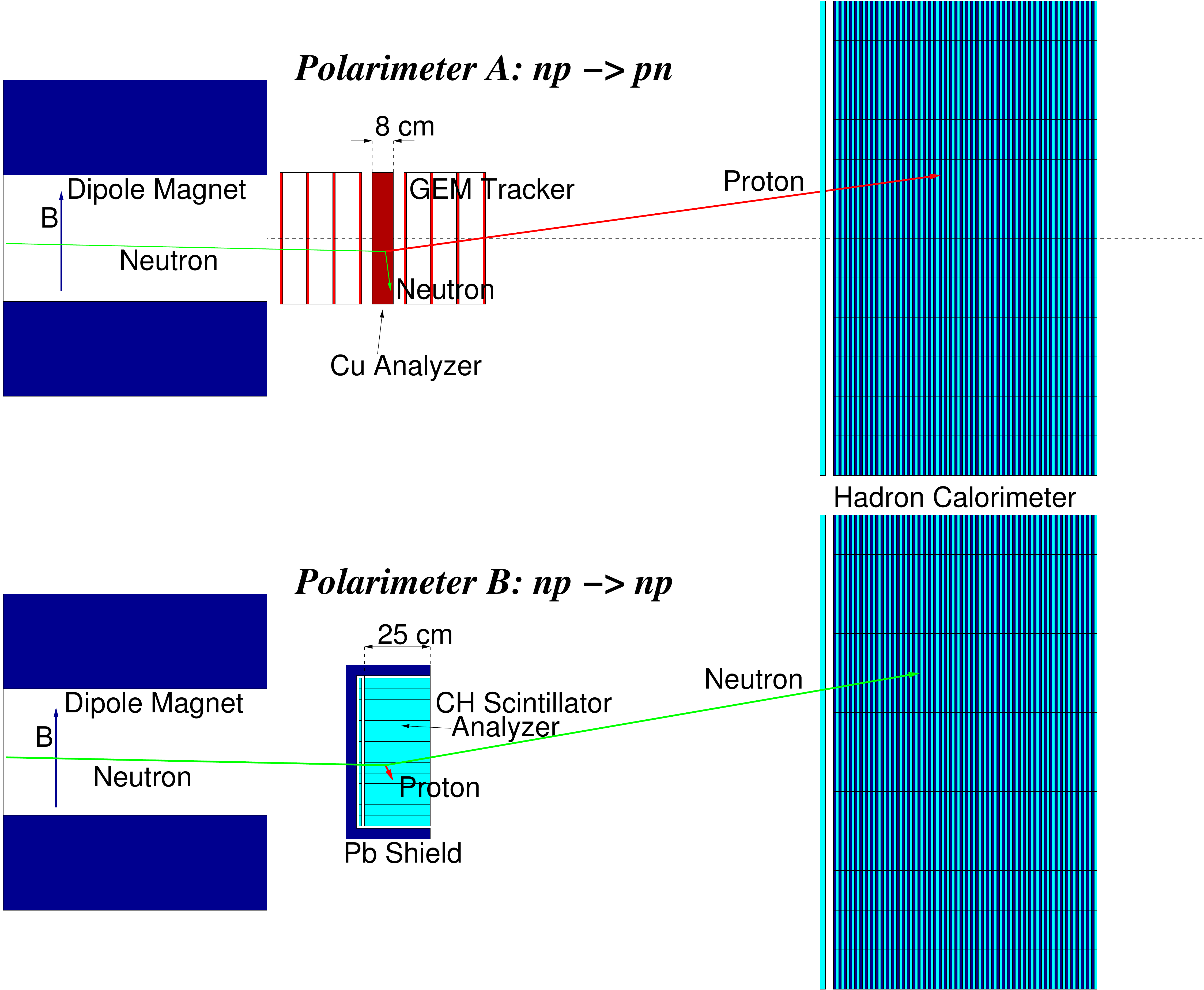}
\caption{Neutron polarimeter configurations considered in the simulations- Polarimeter A for charge exchange reactions (top); Polarimeter B for elastic scattering  (bottom) }
\label{fig:FoMPol}
\end{center}
\end{figure}

\begin{figure}
\includegraphics[width=0.8\columnwidth]{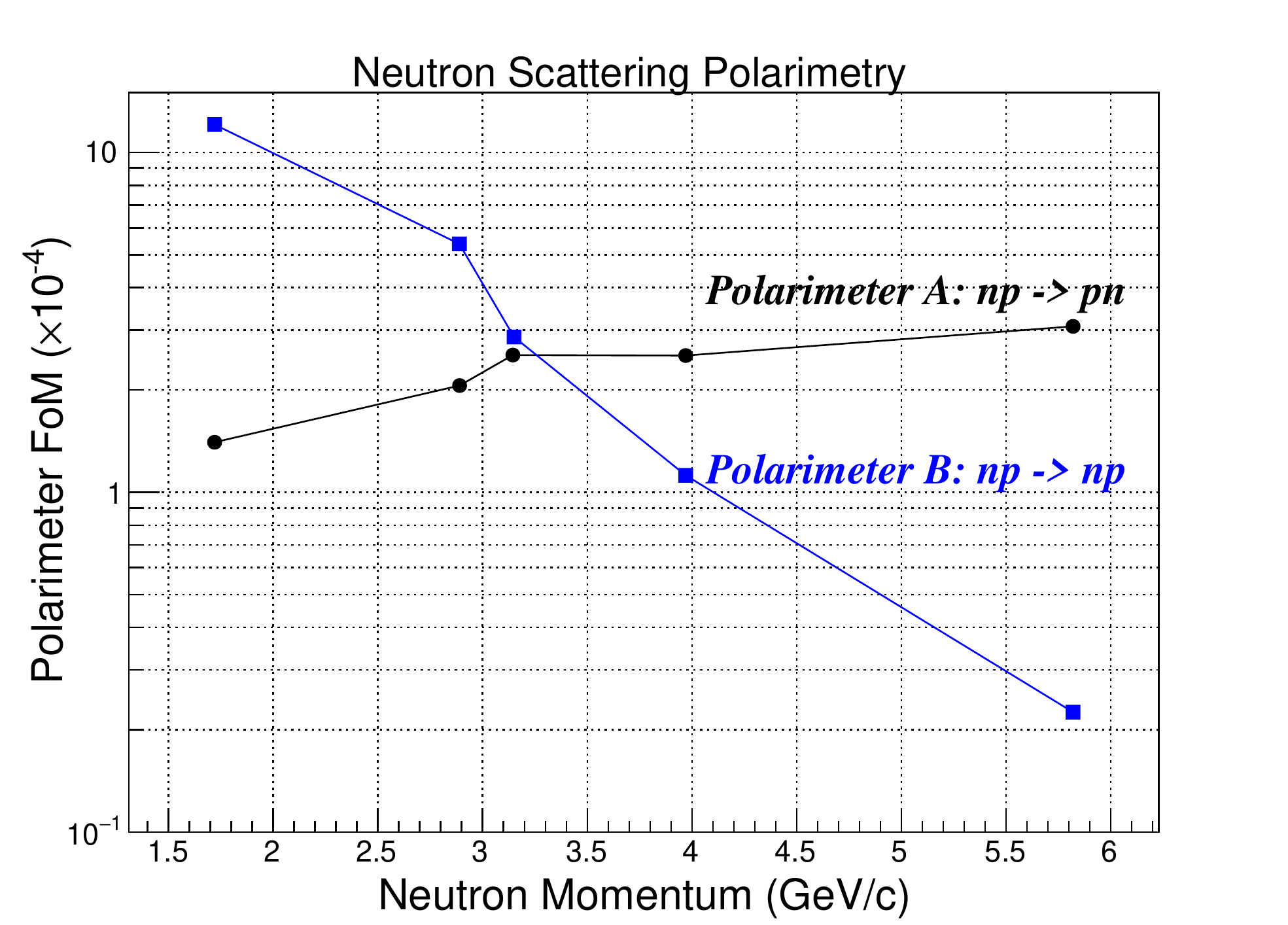}
\caption{Neutron polarimeter figure of merit as a function
of incident neutron momentum for the two polarimeter configurations of Fig.~\protect\ref{fig:FoMPol}.
Blue squares: standard $np$ scattering from $CH$ scintillator (Polarimeter B), black
circles: charge-exchange $np$ scattering from $Cu$ (Polarimeter A).}
%%The red arrow marks the neutron momentum at which a charge-exchange measurement of the analyzing power of Cu was made at Dubna.}
\label{fig:FoM}
\end{figure}

The performance of a polarimeter is usually expressed in terms of 
the figure of merit (FOM),  defined as: 
%\selectlanguage{american}%
\begin{equation}
F^{2}(p_{n})=\int\epsilon(p_{n},\theta_{n}^{'})\,A_{y}^{2}(p_{n},\theta_{n}^{'})d\theta_{n}^{'}
\label{eq:FOM}
\end{equation}
where $p_n$ is the incident momentum, $\theta_{n}^{'}$ is the nucleon scattering angle, $\epsilon$ is the detection efficiency and $A_y$ is the effective analyzing power of the polarimeter. FOM estimates have been calculated over a range of $p_{lab}$ for two polarimeter configurations
(Fig. \ref{fig:FoMPol}) employing either charge-exchange or elastic $np$
scattering and are shown in Fig. \ref{fig:FoM}.
%\selectlanguage{english}%

Elastic-like $pp$ scattering from nuclei is observed to have a factor-two
reduction in $A_{y}$,  compared to the free elastic $pp$. For $np$, the same reduction factor is consistent with the
polarimeter analyzing power obtained in a previous JLab measurement
of $ G_{E}^{n}/G_{M}^{n} $ \cite{PR12-11-009} at 1.45~GeV/c and is assumed for $nC$ scattering.
The value of $A_{y}$ for free elastic $np$ scattering has been
calculated from a fit  to the $pn$ data \cite{Ladygin:409018}. For charge-exchange $np$ scattering from $Cu$, $A_{y}$
was taken from a preliminary analysis of the new data described in
this paper. This analysis has given the dependence of $A_{y}$ on
$p_{t}=p_{lab}\sin\theta_{np}$ at an incident momentum of 3.75~GeV/c.
$A_{y}$ is dependent on $p_{t}$, but has been assumed independent
of $p_{lab}$, in a manner consistent with the free charge-exchange
$np$ data.

Polarimeter efficiencies have been calculated using Geant-4-based Monte Carlo (MC)
simulations of different polarimeter configurations (Fig. \ref{fig:FoMPol}) that record the differential detection
efficiency as a function of scattering angle, after selection of energy and angle ranges in a manner analagous to a real experiment. Calculations have been made for two versions of the polarimeter, compatible
with possible experimental configurations at JLab. 
\begin{enumerate}
\item The polarimeter uses a passive $Cu$ analyzer with forward-angle, charge-exchange
proton detection by GEM trackers and hadron calorimeter (A).
\item The polarimeter employs an active position sensitive $CH$ (plastic-scintillator)
analyser with forward-angle neutron detection by the hadron calorimeter (B).
\end{enumerate}

At neutron momenta above $\sim3.5$~GeV/c, the FOM of charge-exchange
$np$ starts to dominate standard $np$ and by $\sim 6$~GeV/c it
is projected to be a factor $\sim 15$ larger. These calculations suggest strongly that polarimetry by charge-exchange scattering is the technique which will allow recoil polarization  $G^n_E/G^n_M$ to approach $Q^2 \sim$~10~(GeV/c)$^2$.

%%%%%%%%%%%%%%%
\section{The ALPOM2 experiment}
\subsection{The beam production}
%%%%%%%%%%%%%%%

ALPOM2 is placed at the beam line '1v' of the Nuclotron  accelerator facility, at the Veksler Baldin Laboratory for High Energy Physics (VBLHEP) of the Joint Institute for Nuclear Research in Dubna. This beam line was used previously for several experiments using (polarized) neutron beams  \cite{Kirillov:1996zf}, such as the measurement of the $\Delta\sigma(n^\uparrow p^\uparrow)$ total cross section difference \cite{Sharov:2004pg}. 

The polarized deuteron source is provided by the Source of Polarized Ions (SPI), which is a JINR and INR RAS development of the CIPOS source \cite{Agapov:2004uu} formerly employed at the Indiana University Cyclotron Facility. SPI is an atomic beam polarized ion source with a plasma (H,D) charge-exchange ionizer and a storage cell in the ionization region
%%The polarized deuteron beam is provided by the Source of Polarized Ions (SPI), which is an atomic beam polarized ion source with a plasma (H, D)  charge exchange ionizer and a storage cell in the ionization region.  The former polarized source CIPIOS in operation at Bloomington (Indiana, USA) was received after the closing of the Accelerator, and totally renewed at JINR and INR RAS \cite{Agapov:2004uu}.
The ions are  preaccelerated to 100-150 keV in the LU-20 injector, and then accelerated by the Nuclotron \cite{Fimushkin:2016uu}. 

The SPI operates either in polarized or unpolarized mode, following the principle to polarize particles through adiabatic transitions between two hyperfine structure levels, as established by A. Abragam \cite{abragam1961}. 

%\begin{table}
%	\caption{Scheme for the polarization states. } 
%	\label{Table:HFT}
%	\begin{center}
%		\begin{tabular}{|c|c|c|c|c|}
%			\hline
%			HFT1    ?                 & HFT2                         & Final state & $P_z$ & $P_{zz}$  \\
%			\hline
%			MFT3$\to$4?          &WFT 1,2 $\to$ 3,4    & 3,4              & -2/3 &0 \\
%			MFT3$\to$4 ?            & SFT 2 $\to$ 6           & 3,4              & -1 &+1\\	
%			\hline
%			\end{tabular}
%	\end{center}
%\end{table}

The deuteron beam, extracted over a period of ~5 s was incident on a 30(25) cm thick $C(CH_2)$ target  where the deuteron was fragmented into a proton and a neutron. Experiments on polarization transfer from deuteron to proton in break-up reactions showed that the proton (and therefore the neutron) polarization is equal to the polarization of the primary deuteron beam and constant when the internal momenta of the nucleons inside the deuteron $q\le 0.15$ GeV/c \cite{Azhgirey:2008}. Thus, it is possible to produce polarized nucleons with momenta higher than half of the incident deuteron momentum. Deuterons with momentum of 12-13 GeV/c, may produce protons of momentum up to 7.5 GeV/c with a polarization equal to that of the incident deuterons. However, the largest cross section is obtained by selecting secondary nucleons corresponding to  $q$=0. 

In case of neutron beams, protons and deuterons were removed from the neutron beam by a dipole  magnet.  Neutron angles close to zero degrees were selected by a 6 m long collimator made of iron and brass located 13.4 m upstream from the ALPOM2 set-up. The collimators and shielding of the experimental area decreased the low energy tail of the neutron spectrum to about 1\% of 
the peak value.  By setting the primary deuteron beam intensity in the range $ {1-3\times 10^8}$ particles/spill, the average number of protons or neutrons incident on the polarimeter target was kept at the level of $4\times 10^4$ per spill. The length of the spill depends on the primary beam intensity and in the present case was of the order of a few seconds. More details on the operation of the neutron beam line can be found  in Refs. \cite{Kirillov:1996zf,Rukoyatkin:2001nd}.
The polarization of the incident deuterons was oriented downwards, along the vertical axis, perpendicular  to the beam momentum. The polarization of the produced nucleons ($n$ or $p$) keeps the same direction.
%%%%%%%%%%%%%%%%%%%
\subsection{Beam polarimetry}
%%%%%%%%%%%%%%%%%%%
The accurate measurement of the beam polarization is crucial to the extraction of the analyzing power in ALPOM2, as the beam-polarization uncertainty is the main source of the systematic error on the analyzing power. The polarized beam was tagged with its three polarization states, down (plus, '+', $P_z=+1$, $P_{zz}=+1$), up (minus, '-', $P_z=-1/3$, $P_{zz}=0$), and unpolarized (zero, '0'), where the state is changed after each spill. 

The beam polarimeter, denominated F3 as it is located at the focus F3 of the extracted beam line, is based on quasielastic $pp$   scattering, where analyzing powers are known from previous measurements \cite{Bystricky:1981}. F3 has an ionization chamber (IC) as a beam intensity monitor for normalization, and four arms, forward and recoil, left and right. Each arm has three sets of scintillator counters at forward angle $\theta \simeq 8^\circ$,  9$^\circ$,  and 10.5$^\circ$, for momenta 4.2, 3.75 and 3 GeV/c, respectively,  and a bigger scintillator at backward angle $\theta \simeq 60^\circ$ for the recoil particle. The coincidence between forward and recoil arms (left and right) and the IC counts was collected, spill by spill, by the data acquisition system.

The beam polarization was constantly monitored and the stability of the beam was excellent, with polarimeter asymmetry fluctuations not exceeding $2\%$. 
%%%%%%%%%%%%%%%%
\subsection{The ALPOM2 setup}
%%%%%%%%%%%%%%%%

\begin{figure*}
\centering
\includegraphics[width=0.75\textwidth]{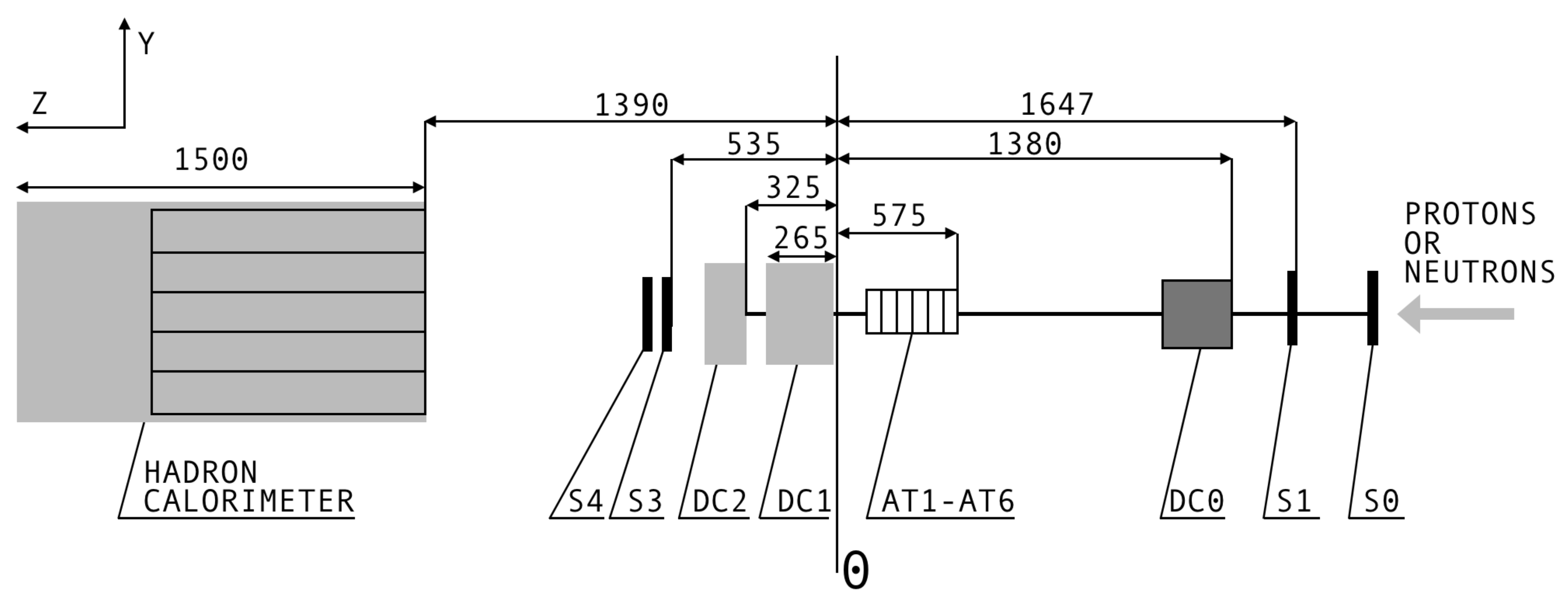}
\caption{Side view schematic of the ALPOM2 set up positioned on the secondary proton/neutron beam line, including: scintillation counters (S0, S1, S3, S4); drift chambers (DC0, DC1, DC2); Hadron Calorimeter. The analyzing targets were located between DC0 and DC1. Here a $CH$ active target (AT1\,-\,AT6), is shown as an example. Dimensions are in mm. $\Theta$ indicates the origin of the $z$ coordinate.}
\label{fig:ALPOM2SetUp}
\end{figure*}

ALPOM2 represents an upgrade of the ALPOM polarimeter \cite{Azhgirey:2004yk}, which in turn is based on the POMME polarimeter employed at Saturne  \cite{Bonin:1989tg}.
A schematic view of the ALPOM2 geometry is shown in Figure~\ref{fig:ALPOM2SetUp}, with the proton/neutron beam travelling along the z-axis, in the longitudinal direction. 
% Image modified from that found in Dima's presentation at Orsay
% Must be replaced with a updated diagram.
% Or if keeping this image, labelled dimensions must be checked by someone.
% It could be nice to add a photo too - request for good quality photograph
The main components consisted of:
\begin{itemize}
\item fast plastic scintillator counters (S0, S1, S3, S4 and optionally AT1-AT6) for triggering purposes;
\item drift chambers (DC0, DC1, DC2) for charged particle tracking;
\item a segmented hadron calorimeter for energy and position measurements of the outgoing particles;
\item the polarimeter analyzer ($C$, $CH$, $CH_2$, $Cu$), where Fig. \ref{fig:ALPOM2SetUp} shows the segmented $CH$ scintillator analyzer (AT1-AT6). 
\item two neutron beam monitors located after the collimator are not shown \cite{Lehar:1996cr}.
\end{itemize}

Several different materials were tested as polarimeter targets (Table \ref{Table:targets}), to measure and compare their analyzing powers.  These analyzers included $C$, $CH_2$, $Cu$ as well as a $CH$ active target with six elements (AT1\,-\,AT6). The downstream face of the analyzer block was located 10 cm before DC1. 
\begin{table}
	\caption{Target material, density and length } 
	\label{Table:targets}
	\begin{center}
		\begin{tabular}{|c|c|c|c|c|}
			\hline\hline
			Target                         & $CH_2$   &  $CH$      & $C$      & $Cu$ \\	
                     \hline 	                   
			 Density [g/cm$^3$]   & 0.919  & 1.06    & 1.68 & 8.96 \\
                     Length [cm]      & 30 & 30 & 20 & 4  \\
			\hline\hline
			\end{tabular}
	\end{center}
\end{table}
The energy deposit in the hadron calorimeter was used to select high energy particles from the elastic-like events of interest, for example 
$\vec p+ CH_{2}\to p+X$, thus reducing contamination from inelastic scattering which produces lower-energy final-state hadrons. In addition it gave position information, which could be correlated with drift-chamber tracks.

The proton polarimeter used S0 and S1 to provide a trigger signal; drift chambers DC0, DC1 and DC2; C, $CH_2$ and $Cu$ targets; the hadron calorimeter; while the neutron polarimeter used S3 and S4 to provide a trigger; drift chambers DC1 and DC2; $CH$, $CH_2$ and $Cu$ targets; the hadron calorimeter. Where the active $ CH$-scintillator analyzer was employed, it was also included in the trigger system. Further details on each component of the set up are given below.

The scintillation counters were used to generate the trigger for the readout of data from all detectors in the set up. Coincident hits from S0  (dimensions $200\times 200 \times 10$ cm) and S1 (dimensions $73 \times 73 \times 5$ mm) were used to trigger on incident protons, during proton polarimetry measurements, whereas S3 and S4 (dimensions $240\times 240\times 20$ mm), located downstream from the target were used in coincidence for triggering during neutron polarimetry measurements. 
%Trigger rates during the proton and neutron runs were typically \textbf{CHECK} and \textbf{CHECK} respectively, with trigger efficiencies on the order of \textbf{CHECK} recorded throughout the data-taking.

The drift chambers were used to reconstruct primary and secondary charged particle tracks  and are described in detail in Ref. \cite{Strela:2013}. Here we summarize their properties. All chambers contain an  ($Ar_2CH_4$) 
%Ar (xx%) CH4 (yy%) 
gas mixture and have alternating, orthogonal X and Y coordinate planes. DC0, of dimensions $12.5\times12.5$ cm, was located upstream from the target and has 8 planes (4X, 4Y). DC1 and DC2 have dimensions $25\times 25$ cm and were located downstream of the target. DC1 has 8 planes (4X,4Y) while DC2 has 4 planes (2X,2Y). 
Corrections for misalignments were performed prior to track reconstructions. To reconstruct a track, hits in at least three chamber planes were required. The hit-position uncertainty provided by the chambers was $<100$ $\mu$m, which produced an angular resolution better than 0.4 mrad, and the reconstruction efficiency for charged tracks was close to 
100\%. 

With proton beams, primary and scattered charged particle tracks were reconstructed using all drift chambers, while for neutron measurements primary signals were absent in DC0. Both protons and neutrons, incident on a nucleus, can produce multiple, charged, final-state particles and indeed some of the registered events have two or more hits in DC1 and DC2.
%%Some of the registered events had two or more hits in the drift planes of chambers DC1 and DC2, due to multi-charged-particle final states. Both protons and neutrons incident on a nucleus could produce multiple charged particles.
A "clean" neutron charge exchange produces one energetic forward proton.

\begin{figure*}
\centering
\includegraphics[width=0.45\textwidth]{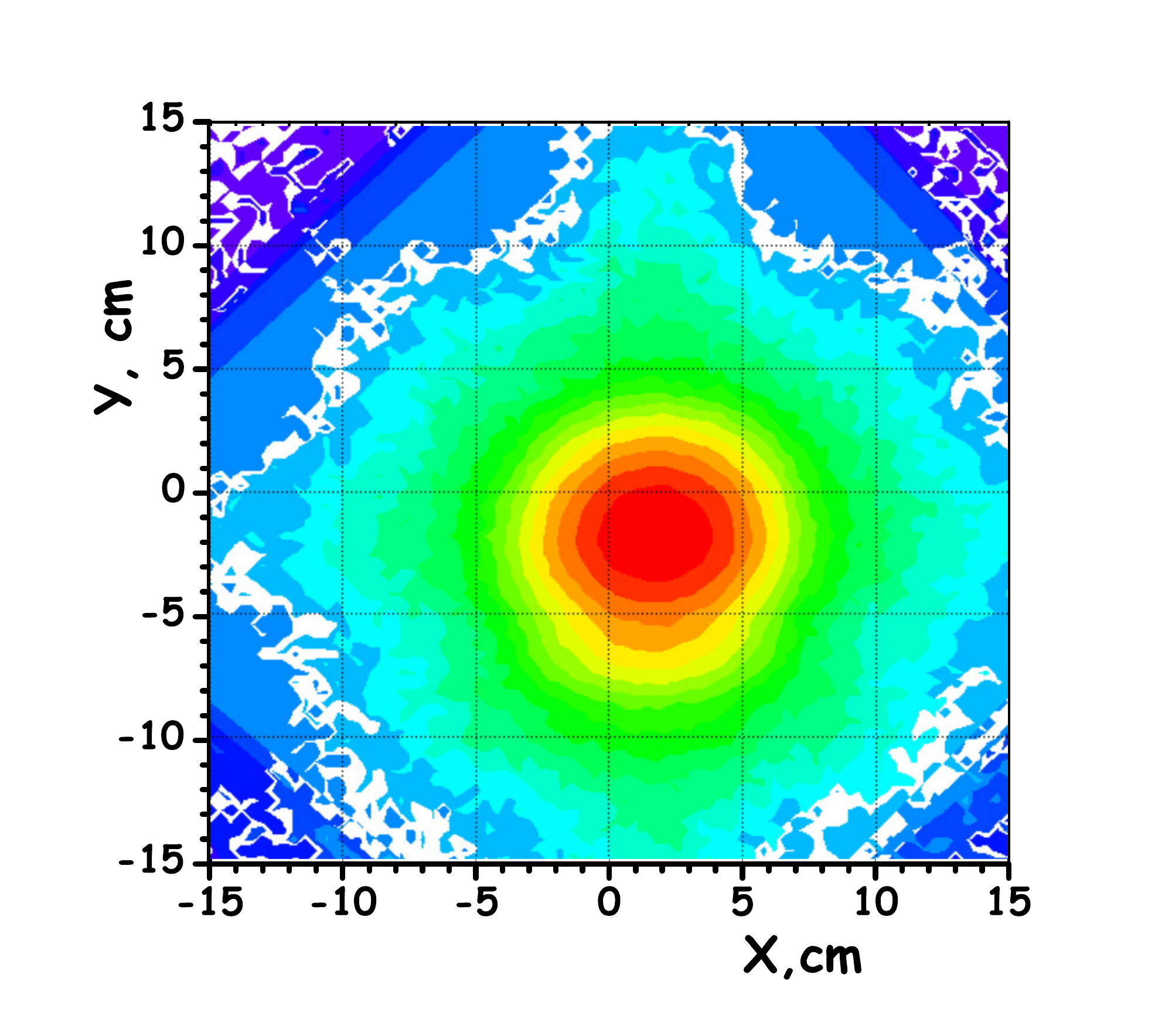}
\includegraphics[width=0.45\textwidth]{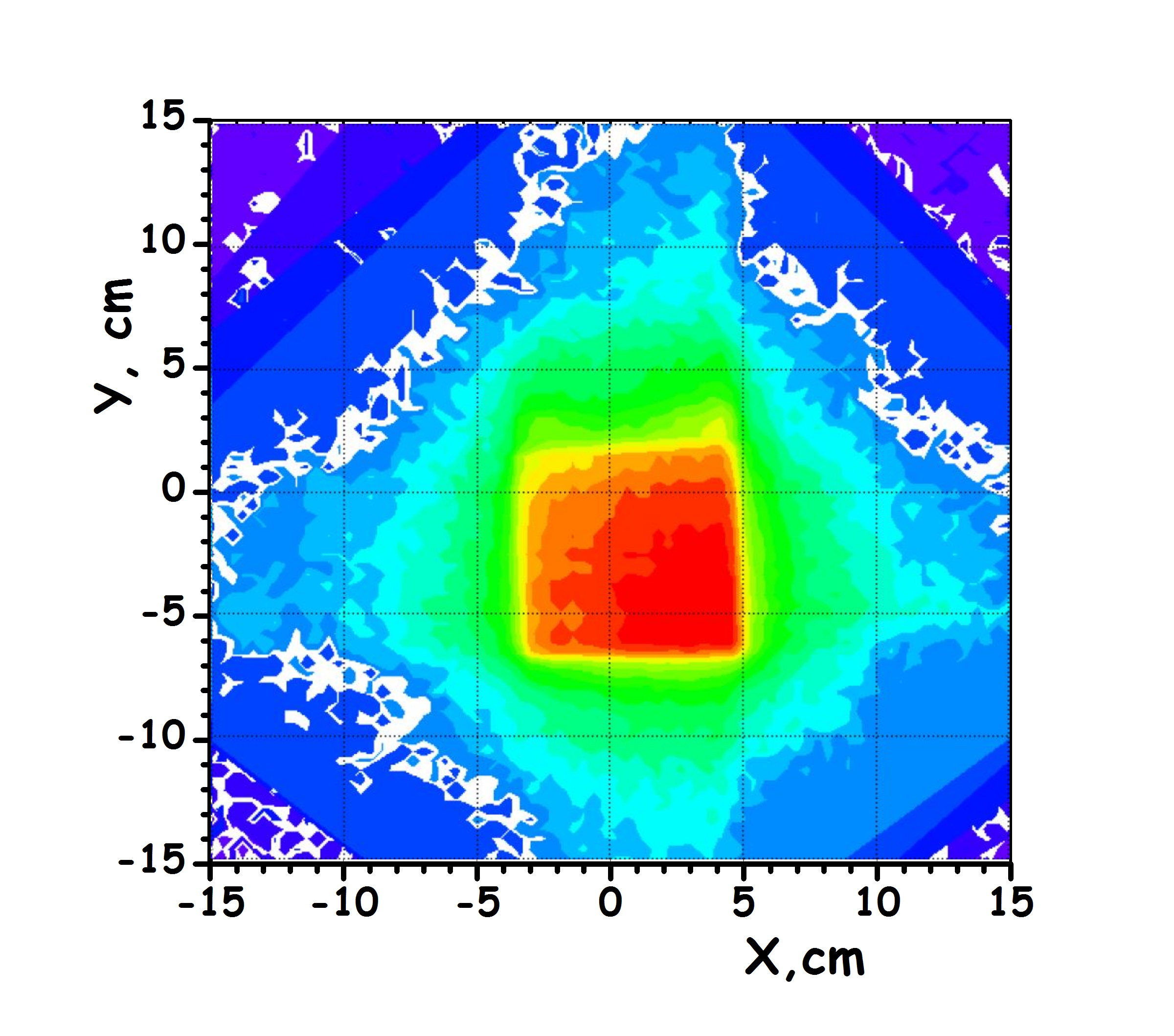}
\caption{Neutron (left) and proton (right) beam profiles in the transverse X-Y plane, as recorded by the drift chambers, for 3.75\,GeV/c momentum and before a 30\,cm long CH$_{2}$ target. }
\label{fig:BeamProfilesDriftChambers}
\end{figure*}

 Examples of the proton and neutron beam profiles, as reconstructed by the chambers are shown in Fig.~\ref{fig:BeamProfilesDriftChambers}. The profile is given by the drift chambers that detect directly the proton beam, while, in the case of  neutron beam, drift chambers detect charged particles generated  from the interaction  of the neutron beam in the target. 

\begin{figure*}
\centering
\includegraphics[width=0.45\textwidth]{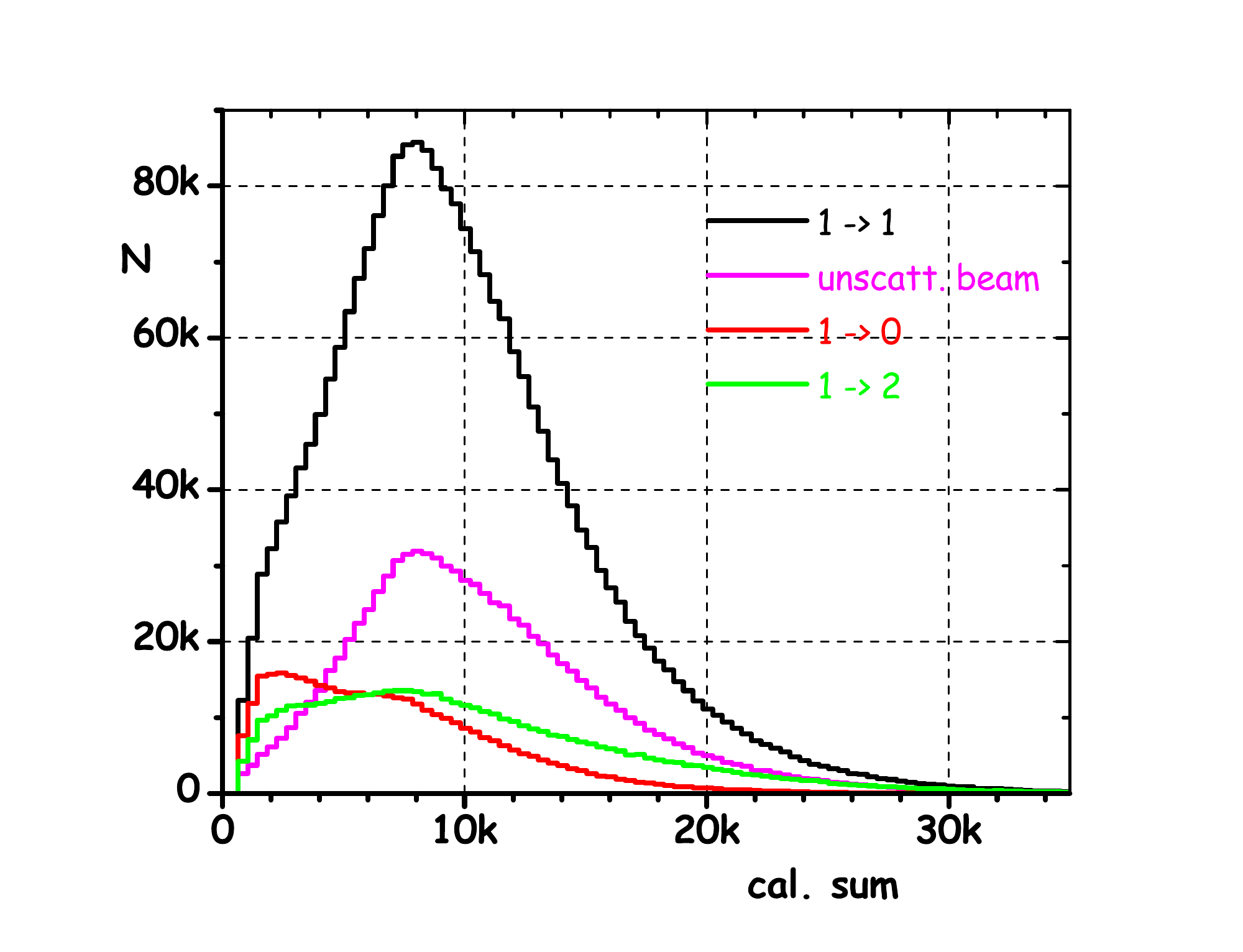}
\includegraphics[width=0.45\textwidth]{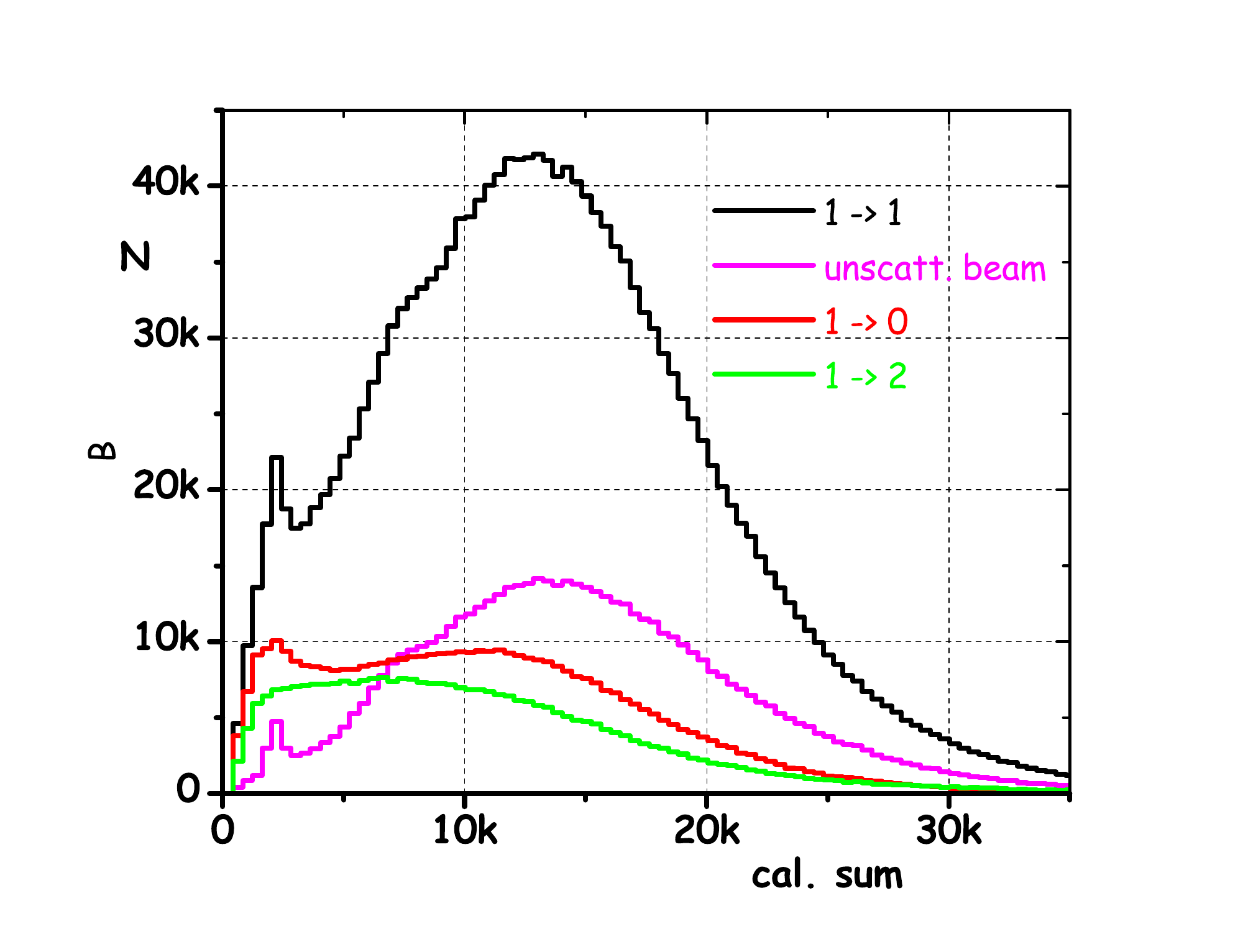}
\hspace*{-0.1cm}
\caption{Summed calorimeter energy deposit for different processes induced on a $CH_2$ target by protons of  momentum 3.75 GeV/c (left) and 6.0 GeV/c (right).}
\label{p_process}
\end{figure*}

\begin{figure*}
\centering
\includegraphics[width=0.65\textwidth]{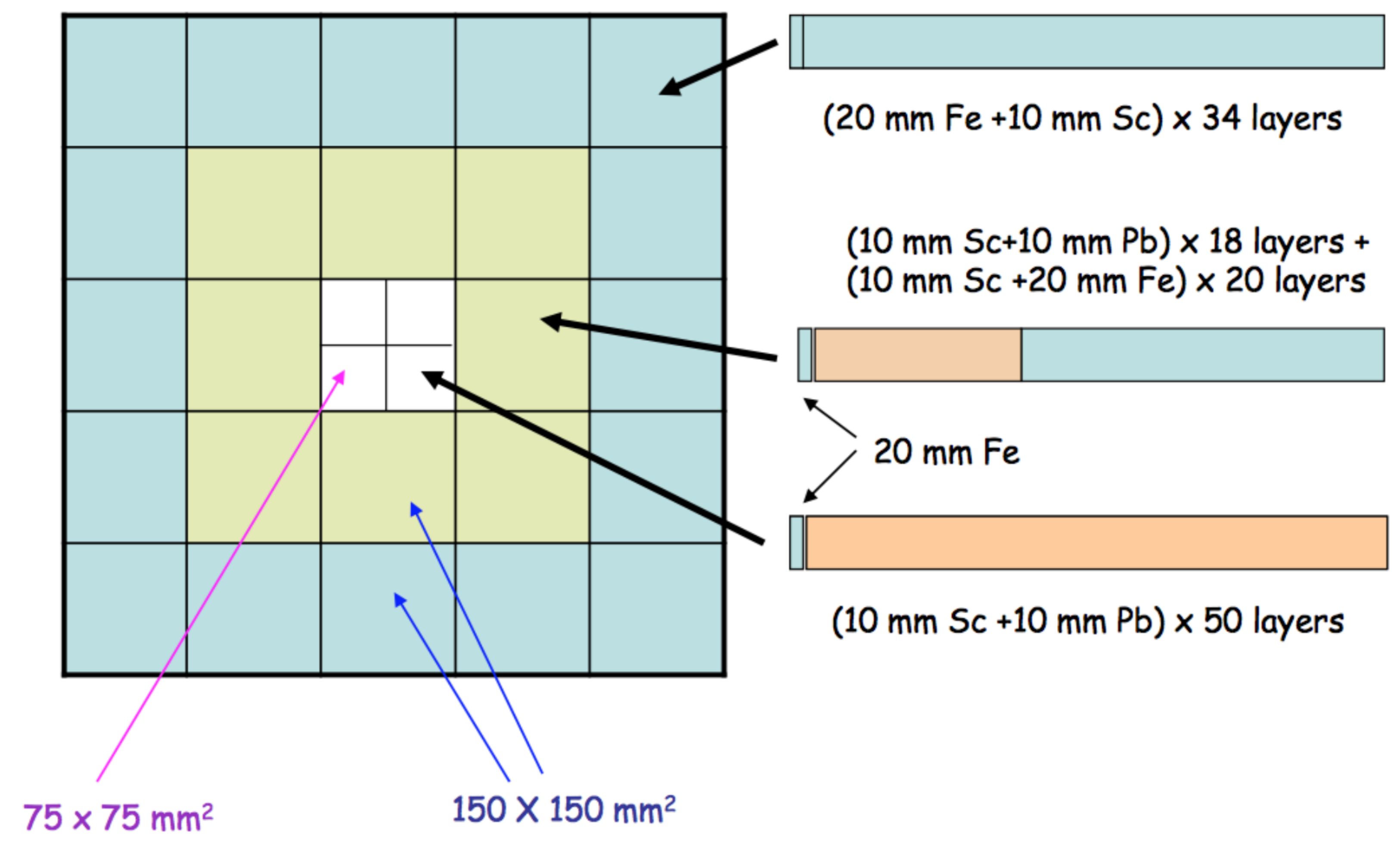}
\caption{View of the different bars of the hadron calorimeter used in the ALPOM2 set up. The different bar compositions are noted at the side.}
\label{fig:HadCalDesign}
\end{figure*}

The hadron calorimeter, located at the downstream end of the polarimeter increased the fraction of useful, elastic-like events selected. Examples of the hadron calorimeter response, from protons incident on a $CH_2$ target,  are given in Fig. \ref{p_process}. The response has been sorted into four categories: no interaction with the target ($NI$); no final-state charged particle ($1 \to 0$); one final-state charged particle ($1 \to 1$); two final-state charged particles  ($1 \to 2$). The  $1 \to 1$ category is the selected one to obtain the analyzing power.

It is separated from the $NI$ category on the basis of the angle between incoming and outgoing tracks, which is chosen to be $>1.7^\circ$ to exclude multiple Coulomb scattering.

In the case of a neutron beam making a charge exchange in the target, the events of interest are no incident charged and one final-state charged particle ($0 \to 1$).

When the proton beam momentum increases from 3.75 GeV$/c$ to 6 GeV$/c$, the proportion of  $1 \to 1$ events decreases dramatically. Other processes, such as $NI$ or $1 \to 2$, are enhanced relative to $1 \to 1$ which decreases the scattering asymmetry if they are not excluded. Neutron induced reactions are expected to have a similar behavior when the beam momentum increases. Placing a threshold on the summed energy deposit of the hadron calorimeter helps to identify the useful reactions. This is shown in Fig. \ref{p_process}, where the number of events corresponding to each type is shown as a function of the sum of the energy deposited in the calorimeter, at 3.75 GeV/c (left) and 6.0 GeV/c (right) for a $CH_2$ target.
%{\color{red} need better description} {\color{green} hope it is clear now.}
%
%%%%%%%%%%%%%%%%%%
%% Magenta should be labeled ``NI'' for consistency with text
%% I do not find the figure totally convincing. Maybe a small table giving integrated counts: 1) no calsum threshold,
%% 2) above a calsum threshold e.g 1/2 peak amplitude (showing threshold on plots) for both 3.75 and 6 GeV/c would quantify somewhat ...jrma
%%%%%%%%%%%%%%%%%%%%%
%
The ability of the hadron calorimeter to select high-energy nucleons will be extremely important to the forthcoming  JLab EMFF experiments \cite{PR12-17-004,PR12-07-109}, which will operate at very high luminosity, where the electron beam will generate a huge flux of low-momentum background. The hadron calorimeter will provide the primary trigger for the nucleon arm of these experiments and the selection of high momentum particles through their energy deposit will be vital for suppressing soft background in the polarimeter. Furthermore it will help to select nucleons produced at forward angles where the analyzing power will be largest.

%%Such study is particularly important for the forthcoming JLab experiments. The new large acceptance spectrometer built for the future GEp-V experiment at high transfer momenta (up to 12 GeV$^{2}$) will have extremely large particle rates \cite{PR12-07-109}. 

%%Moreover, as the analyzing power becomes smaller with increasing momentum, the calorimeter can be used to select the leading protons at smallest angles, which has the effect to increase the analyzing power. The hadron calorimeter makes it possible to select high momentum particles through their energy deposit. 

The hadron calorimeter, built in Dubna for the COMPASS experiment at CERN  \cite{Vlasov:2006,abbon:2007},  is a sampling calorimeter composed of alternating plates of plastic scintillator and iron or lead. The iron/lead provides most of the stopping power for energetic hadrons, so that the incident energy is totally absorbed in a thickness  less than 1 m. As used in ALPOM2, it comprises  28 elements (Fig.~\ref{fig:HadCalDesign}). Four bars with dimensions $75\times75$\,mm, were used in the central region, at smaller scattering angles and where higher count rates were experienced,  whereas at larger scattering angles, 24 bars with dimensions of $150\times150$\,mm were used.

\begin{figure}[h]
\centering
\includegraphics[width=8.5cm]{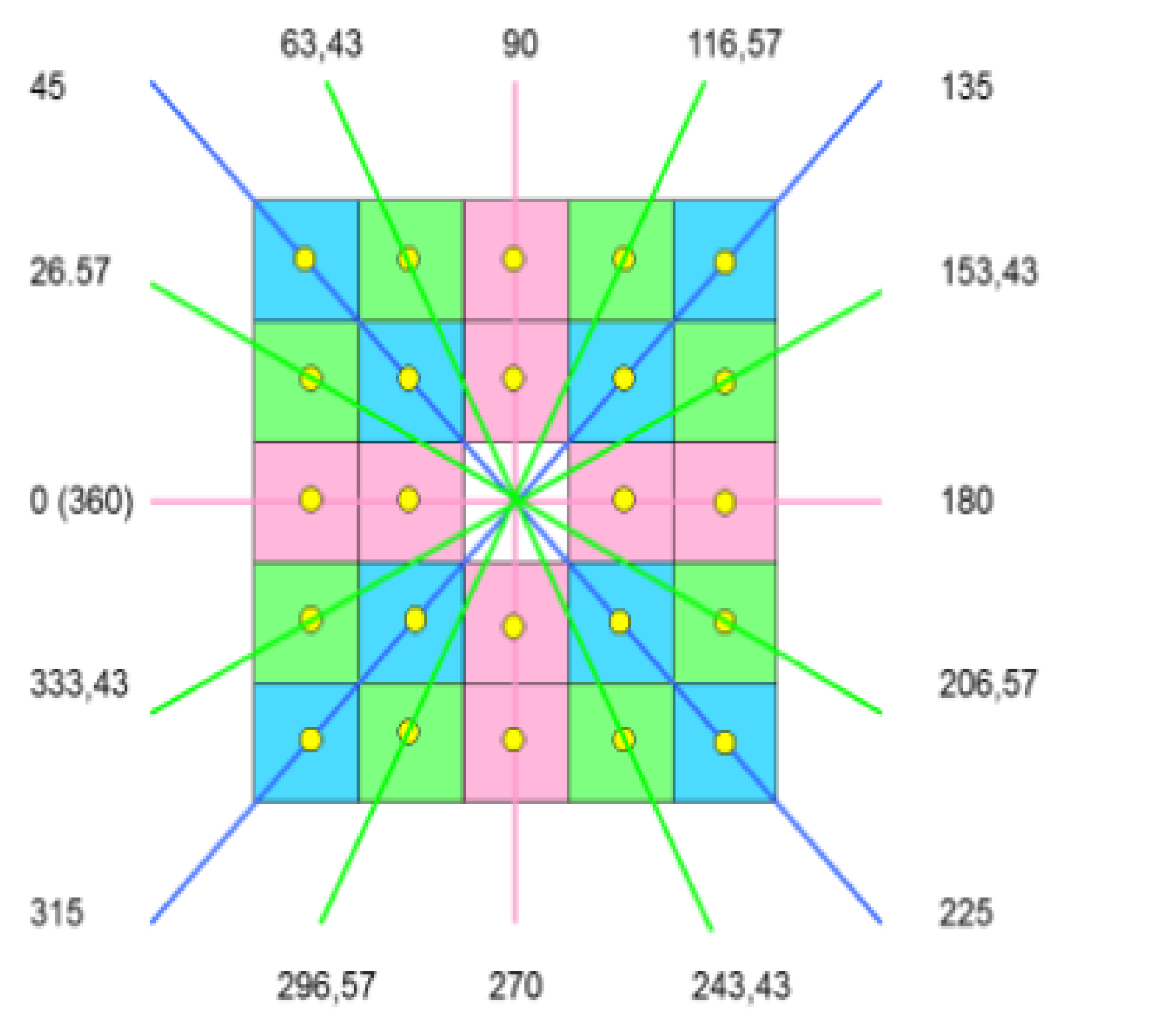}
\caption{Azimuthal angles corresponding to the central points of the hadron calorimeter bars.}
\label{fig:HadCalAng}
\end{figure}

As shown in Fig.~\ref{fig:HadCalDesign}, different bars have different arrangements of iron (Fe), lead (Pb) and scintillator (Sc) that was 10 mm thicker in comparison with the original COMPASS bar. The azimuthal angle selection, determined by the segmentation of the calorimeter and subsequently used in presentations of asymmetries, is given in Fig.~\ref{fig:HadCalAng}.

\begin{figure*}
 \centering
\includegraphics[width=0.75\textwidth] {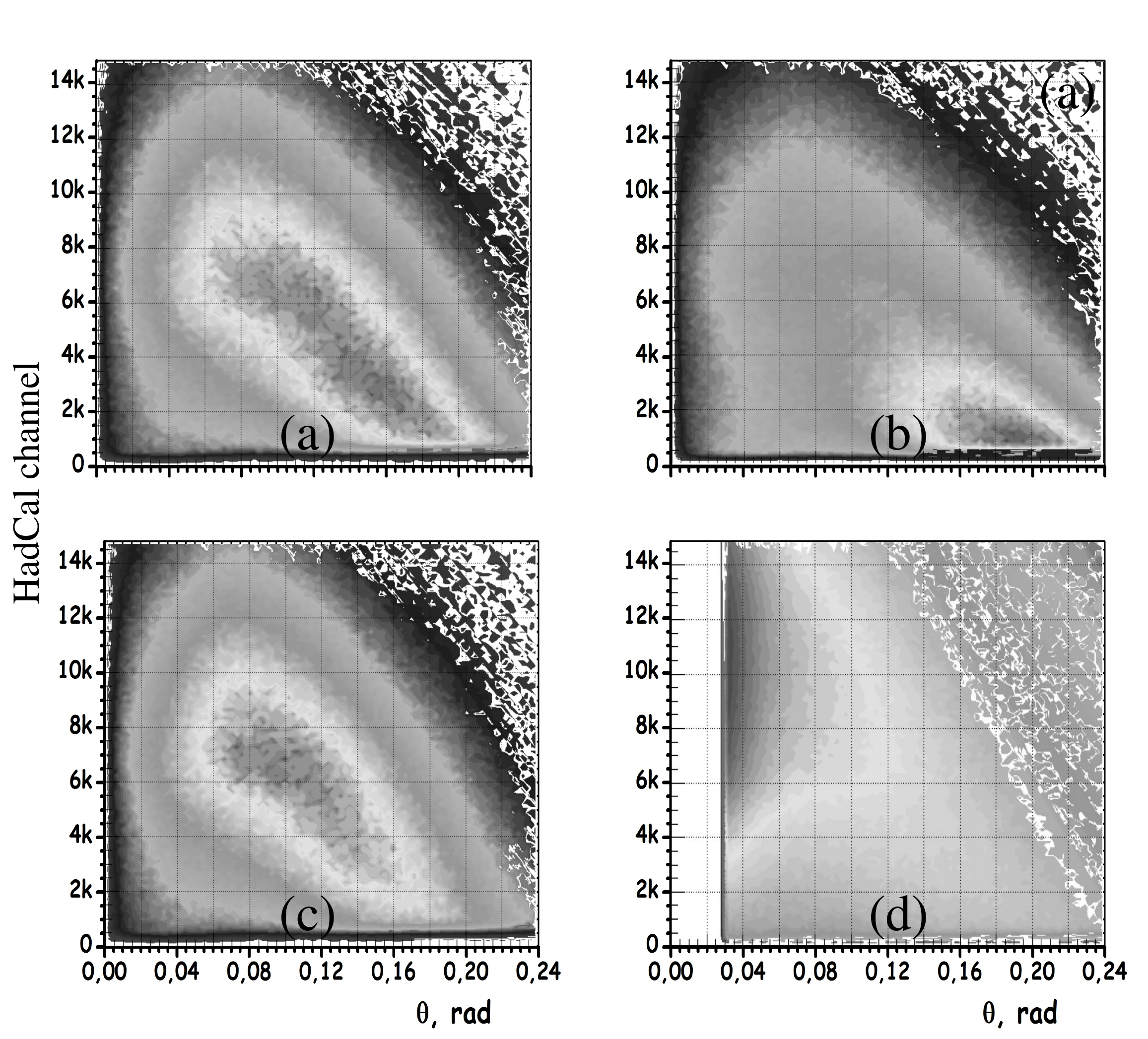} 
\caption{Hadron calorimeter summed energy deposit vs. particle angle 
for a) $n +C$, b) $n +Cu$, c) $n +CH_2$, and d) $p +CH_2$. In subfigure d) the events corresponding to the 
unscattered beam are removed by a small angle cut. }
 \label{Fig:hadcal-response} 
\end{figure*}

The response of all calorimeter bars and their associated electronics was calibrated in dedicated cosmic-ray runs, where the hadron calorimeter was rotated by 90 degrees so that the bars were aligned vertically. A further calibration, with the calorimeter in standard alignment, were performed with the proton beam.  The results are shown in Table \ref{Tab:hadro} and in Fig. \ref{Fig:hadcal-response}.
 \begin{table}
	\caption{The peak energy deposit (channel) and the peak width (FWHM) are given for a proton beam on a $CH_2$ target and for a neutron beam on an active target (CH), after the hadron calorimeter calibration with cosmic rays.} 
	\label{Tab:hadro}
	\begin{center}
		\begin{tabular}{|c|c|c|c|c|}
			\hline\hline
			Particle         & Momentum      & T           &  Energy    & Width  \\
			                    & [GeV/c]            &  [GeV]   & [channel]  & [channel]  \\
			\hline
			proton         &3.00    & 2.205   & 7300   & 3400 \\
			proton         &4.20    & 3.365   & 11100  & 4700 \\
			proton         &3.75    & 2.927   & 10000  & 4500 \\
			neutron       &3.75    & 2.926   & 9700    & 4600 \\
			\hline\hline
			\end{tabular}
	\end{center}
\end{table}

%%%%%%%%%%%%%%%%%%%%%%%%%
\subsection{Readout and Data Acquisition}
%%%%%%%%%%%%%%%%%%%%%%%%%%
Signals from the detectors were first shaped and processed by a combination of custom made front-end electronics, and subsequently readout via a VMEbus data acquisition (DAQ) system. The DAQ controlled data readout from several VMEbus modules, including a trigger control module, a multi-hit scaler, multi-hit time-to-digital converters (TDCs), and time and charge waveform sampling digitizers (TQDCs). The DAQ electronics, online software and modules were developed by the AFI electronic group of JINR \cite{Strela:2013}.

The TDCs were used to record signal times from the drift chambers, scintillation counters and a flag giving  the polarisation state of the deuteron beam, whereas, the TQDCs were used  to record signal timing and amplitude from the hadron calorimeter and active $CH$ analyzer. Time resolution of 100\,ps was provided by the TDC and TQDC modules.

%%%%%%%%%%%%
\section{Data Analysis and  results}
%%%%%%%%%%%%%%%%}

\begin{figure}[h]
 \centering
\includegraphics [width=0.45\textwidth] {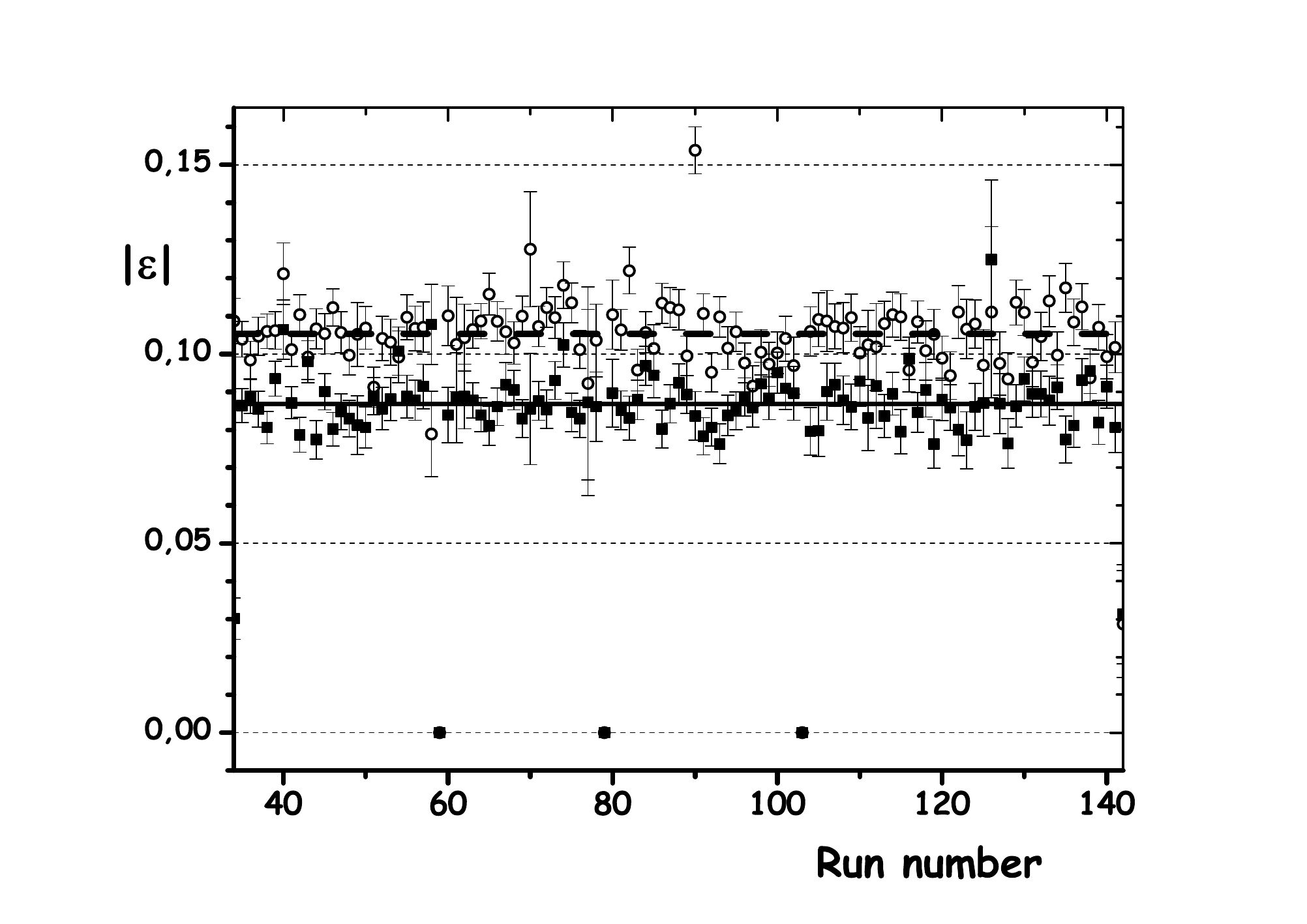}
\caption{Left-right asymmetry from F3,  for each polarization state as a function of the run number during the data taking.}
%%The counting rate, burst by burst for the $+$ and $-$ states is normalized to the conting rate in the unpolarized state}
\label{polarim}
\end{figure}

In Fig.  \ref{polarim}, the left-right asymmetries from the F3 polarimeter for each state of beam polarization
%%the number of events for each polarization state normalized to the unpolarized state
are displayed as a function of the run number, together with lines denoting the mean values. This demonstrates the excellent stability of the beam polarization throughout the data taking, as well as a small difference between the asymmetries obtained in the different  polarization modes. 

The results on analyzing powers given here are based on the following values of the
polarization: 

$(P^{-}, P^{+})=(0.302,0.590) $ for the runs in November  2016 and  

$(P^{-}, P^{+})=(0.434,0.525)$ for the runs in February 2017.

The analysis procedure is described as follows. Ingoing and outgoing trajectories were reconstructed from the drift chambers, the azimuthal and polar angles calculated and their histograms built for each polarization state of the beam.  The results are plotted as functions of the transverse momentum $p_t=p_{lab}\sin \theta$, that may be related approximately to  the Mandelstam variable $t$, by
$p_t\simeq \sqrt{-t}$. The considered process not being completely elastic, the variable $t$ is not directly related to the momentum transferred to the system.

%%%%%%%%%%%%%%%%%%%%%%%%%
%%\subsection{Cross sections}
\subsection{Nucleon Yield Distributions}
\begin{figure}[h]
 \includegraphics[width=0.49\textwidth] {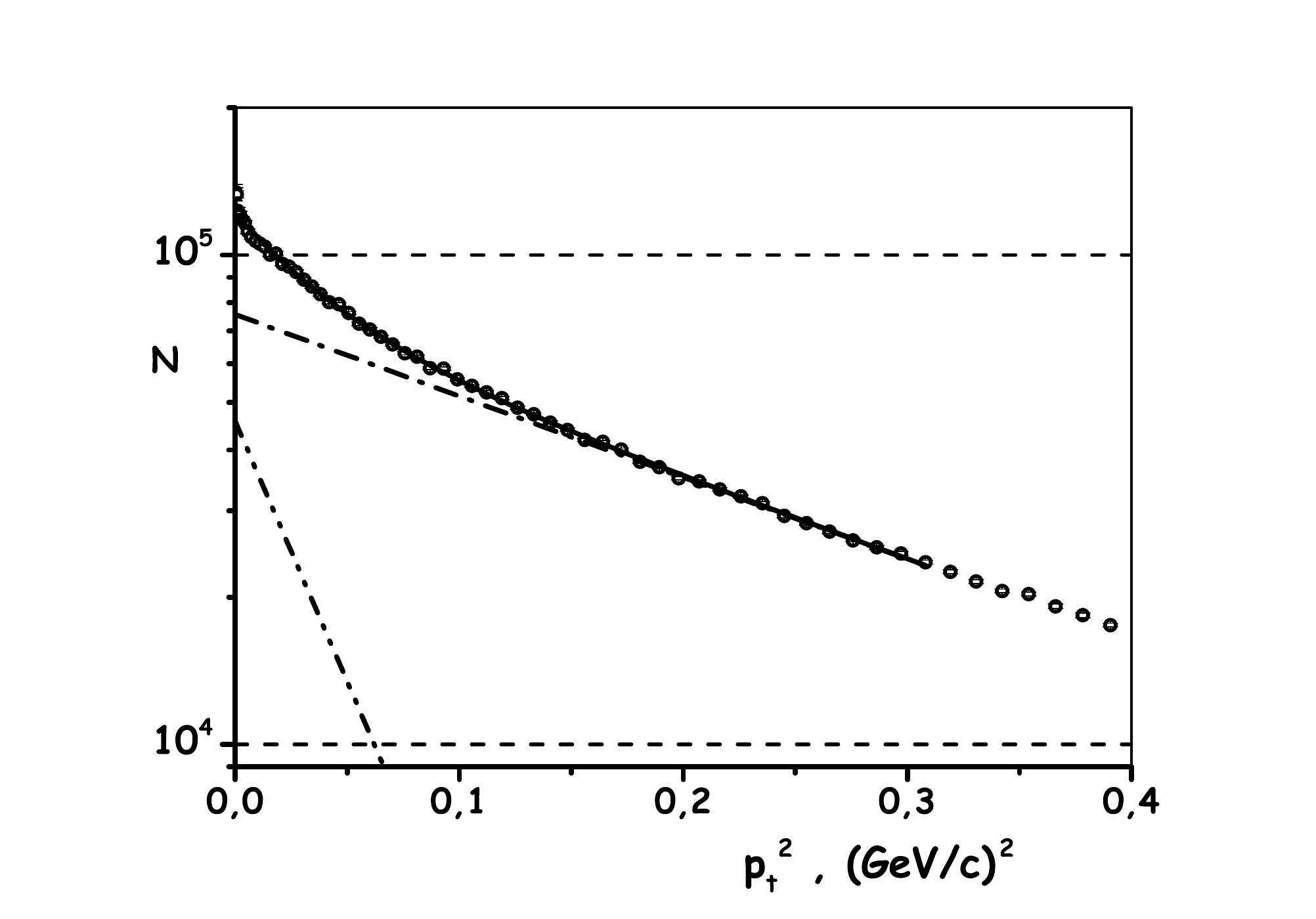} \hfill
\parbox[t]{0.47\textwidth}
{\caption{\small $p_t^2$-distribution for $n+C$ scattering  events at 3.75 GeV/c (arbitrary units).
 The solid line is the sum of  exponential  functions, the dot-dot-dashed and the dot-dashed lines  correspond respectively to the two contributions with slope parameters $b_1$ and $b_2$.}
 \label{Fig:ptn}}
\end{figure}

\begin{figure}[h]
\includegraphics
[width=0.49\textwidth] {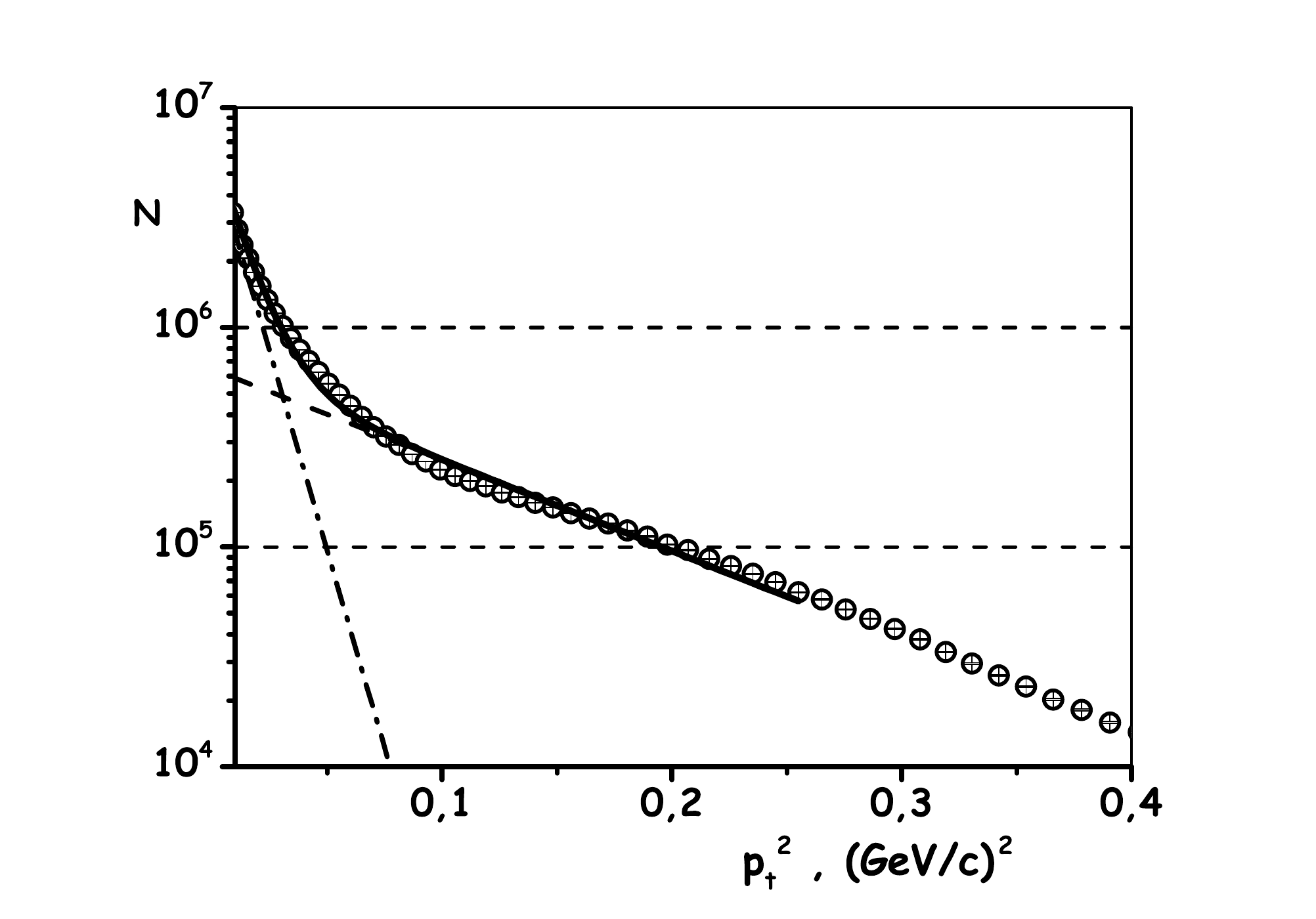}
\parbox[t]{0.47\textwidth}
{\caption{\small $p_t^2$-distribution for $p+CH_2$ scattering events at 3.75 GeV/c  (arbitrary units).
 The solid line is the sum of exponential functions, the long dashed and the dot-dashed lines correspond to the two contributions with slope parameters $b'_2$ and $b'_3$, respectively.}
 \label{Fig:ptp}} 
 \end{figure}

%% the entities displayed are not cross sections
%%%%%%%%%%%%%%%%%%%%%%%%
The nucleon yield distributions as a function of $p_t^2$ are displayed in  Fig. \ref{Fig:ptp}, for $p+CH_2$ scattering and in Fig. \ref{Fig:ptn}, for $n+C$ scattering at momentum 3.75 GeV/c.
The analysis has been limited to the kinematic region $p_t<0.4$ GeV/c in order to avoid the very large corrections where the acceptance of the polarimeter is low. 
These distributions are the convolution of physical processes and finite resolution effects of the detection system.
The  differential cross section for elastic nucleon scattering in the GeV region is in general well described by a sum of  $t$-dependent exponentials:
$$\frac{d\sigma}{dt}=\sum_i c_i \exp(b_it). $$
Thus, in the $p_t^2$ representation,  a satisfactory description of the data in terms of a similar exponential sum is expected:
$$\frac{d\sigma}{dp_t^2}=\sum_i c_i  \exp(-b_ip_t^2). $$
where the parameters $b_i$ and $c_i$ were determined from fits to the distributions of Fig. \ref{Fig:ptn} with the help of the software package FUMILI \cite{Sitnik:2014zwa}.

%%%%%%%%%%%%%%%%%%
%% what does the np show in terms of slope?? otherwise omit...jrma
%%%%%%%%%%%%%%%%%% with increasing energy.
Early studies of $np$ elastic scattering have been made at energies below 6 GeV \cite{Kreisler-SLAC1966,Bystricky:1981} and in the range 5 to 30 GeV \cite{Gibbard:1971rt}. They show that $np$ cross sections are similar to $pp$ cross section, with similar slopes, which implies that  the interaction radii are similar. The $np$ system shows  a shrinkage of the diffraction peak with increasing energy.

The $p_t$-distribution for $n+C$ scattering at 3.75 GeV/c is presented in Fig. \ref{Fig:ptn}.
The distribution is described by the sum of two exponential functions
with slope parameters $b_1=24.5$ (GeV/c)$^{-2}$ and  $b_2=3.2$ (GeV/c)$^{-2}$.
%The data displayed in Figs. \ref{Fig:ptp} and \ref{Fig:ptn} were fitted with the help of FUMILI minimization package \cite{Sitnik:2014zwa}.
To our knowledge there is no previous data to compare with the present $n+C$ slopes.

The $p_t^2$ distribution for $p+CH_2$ scattering at 3.75 GeV/c is described by the sum of three exponential  functions. The first, whose form is related to multiple, small-angle Coulomb scattering convoluted with the experimental angular resolution, is not shown in Fig. \ref{Fig:ptp}. The  slope parameter  for the second function is $b'_2=71.3\pm 0.2 $ (GeV/c)$^{-2}$, which is close to the slope parameter for $p+C$ elastic scattering, previously determined to be $69 \pm 4$ (GeV/c)$^{-2}$ \cite{Jaros:1977it}. The third component has $b'_3=7.4\pm 0.1$ (GeV/c)$^{-2}$, that 
is close to the slope of $pp$ elastic scattering, which varies between 7 and 8.

%%%%%%%%%%%%
%% Moved this to more appropriate place...jrma
%%The kinematical region is limited to $p_t<0.5$ GeV/c in order to avoid large corrections due to the reduced setup acceptance.
%%%%%%%%%%%%%

%%%%%%%%%%%%%%%%%%%%%%%%
%% I don't think the following is suitable for an article in a refereed journal
%% especially if we go for EPJ, which I think fits better than NIM....jrma
%%%%%%%%%%%%%%%%%%%%%%
%%The data displayed in Figs. \ref{Fig:ptp} and \ref{Fig:ptn} were fitted with the help of FUMILI minimization package \cite{Sitnik:2014zwa}. In order to stress the simplicity of the use of this package, the input lines to fit the distribution for the $nA$ interaction are reported here:
%%\begin{verbatim}
%%external fumray
%%np=4  ! number of parameters
%%h=fumifiles('*',np,'ptn.dat',3,100,fumray,30,'ptn.res','tc',1)! sum of 2 exponentials
%%end
%%\end{verbatim}
%%Here $fumray$ is the reference to the user function, 'ptn.dat' is the name of the data file, 'ptn.res' is name of the output file, np=4 is the number of parameters (the distribution is described by the sum of two  functions, with two parameters each (normalization and slope).
%%\end{comment}
%%%%%%%%%%
%asymmetry
%%%%%%%%%%%%%%%%%
%%%%%%%%%%%%%%%%%%%%%%%%%%%%%%
\subsection{Analyzing powers}
%%%%%%%%%%%%%%%%%%%%%%%%
The number of events detected in a unit solid angle around the direction of the scattered particle,  for beam polarization states ``+" and ``-", $N^{\pm}(\theta, \phi)$, is:
\ba
N^{\pm} (\theta, \phi)&=& N_{0}^{\pm}(\theta) (1\pm P^{\pm} A_{y}(\theta)\cos\phi).
\ea
The beam polarization did not have the same value  in the two states, $i.e.,$ 
$P^{+} \neq P^{-} $.  $N_{0}^{\pm}$ is the number of events for unpolarized beam, where the superscript ${\pm}$ indicates  that the beam intensity was different in the two polarization states, requiring a different normalization, $\alpha=N_{0}^-/N_{0}^+$.
Instead of $\theta$, the variable $p_t$ is preferred. 
The asymmetry $\varepsilon$ and its error are derived for each $p_t$ beam, from a linear fit in $\cos\phi $
using  the following formulae:
\be
\varepsilon(p_t,\phi)=  \displaystyle\frac{N^+(p_t,\phi)- \alpha N^-(p_t,\phi)}
{N^+(p_t,\phi)P^- + \alpha N^-(p_t,\phi) P^+ }=A_{y}(p_t) \cos\phi . 
\label{Eq:Asym}
\ee
In this way we also cancel experimental asymmetries, due, for example, to apparatus or beam misalignment, that are present for all beam polarization states. The related statistical error is derived by propagating the errors on $N^\pm$ and $P^\pm$. 

The main source of systematic error on the analyzing power comes from the precision with
which the beam polarization is known.

\begin{figure}[htp!]
\includegraphics[width=7.5cm]{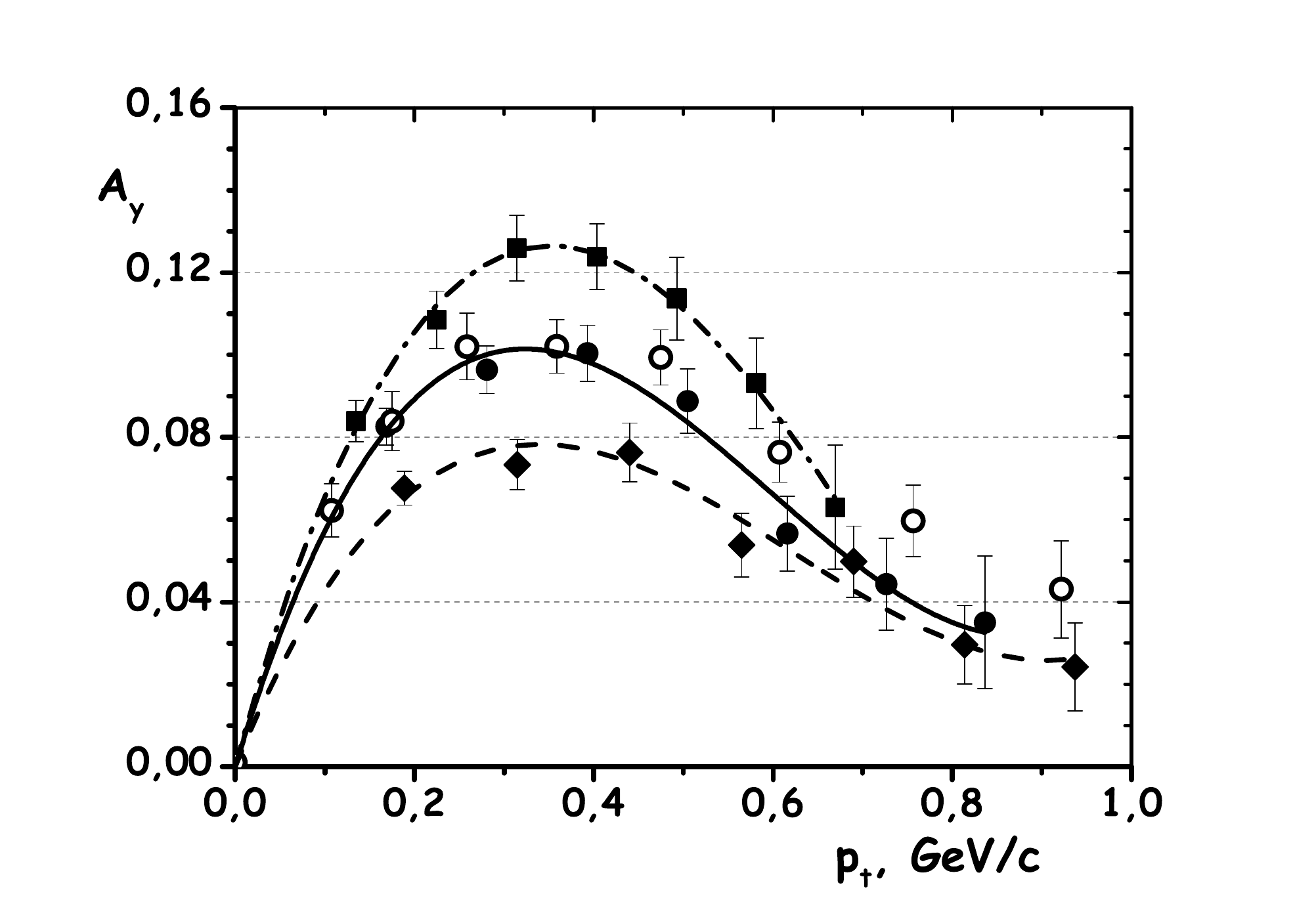}
\caption{$A_y$ for $p+CH_2\to p+X$ at beam momentum 3.75 GeV/c (solid circles), measured in this experiment, compared to  previous data from Ref. \cite{Azhgirey:2004yk} at beam momentum 3.0 GeV/c (solid squares), 3.8 GeV/c (open circles)  and 4.2 GeV/c (solid  lozenges). The lines represent eye guides through the data points.}
\label{Fig:p+CH2_3-42}
\end{figure}

The present analyzing powers for $p+CH_2$ at 3.75 GeV/c momentum (solid circles in Fig. \ref{Fig:p+CH2_3-42}) are consistent with previous data obtained at 3, 3.8 and 4.2 GeV/c  \cite{Azhgirey:2004yk}. All displayed data follow a similar $p_t$ dependence and show a systematic decrease of $A_y$ with increasing incident momentum.

\begin{figure}[htp!]
\includegraphics[width=7.5cm]{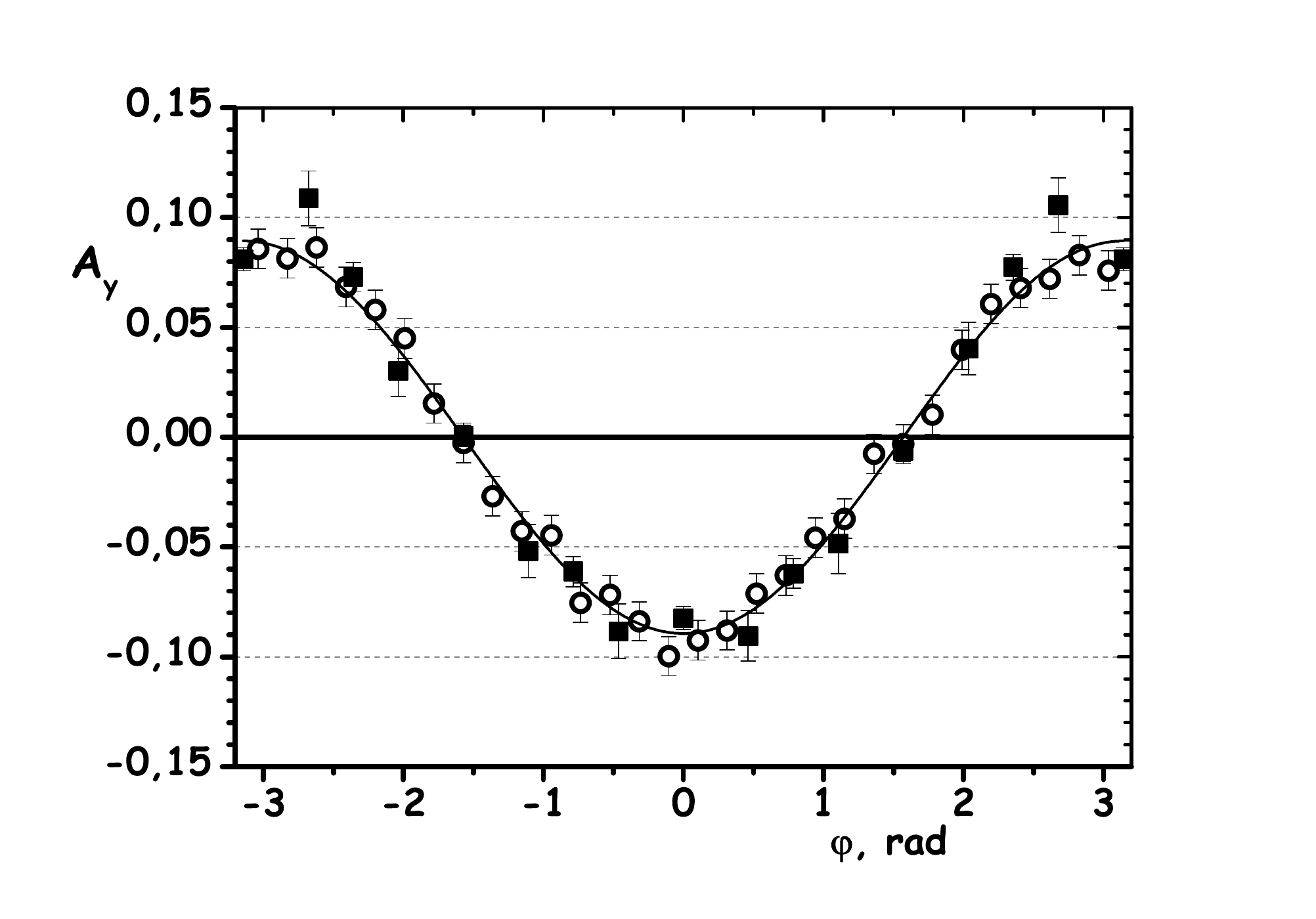}
\caption{$A_y$ for $p+CH_2$ scattering at a momentum 3.0 GeV/c determined from the hadron calorimeter (filled black squares), compared to averaged values determined from the tracks (empty circles). }
\label{Fig:tracks-hadcal}
\end{figure}

The scattering asymmetry can be derived independently from the angle information given both by DC1, DC2 and by the hadron calorimeter.  In order to check the reliability of the analysis based on drift chambers we also reconstructed the asymmetry from the hadron calorimeter, without the central four small bars,  although with coarser granularity in the azimuthal angle. The results for $p+CH_2$ at 3.0 GeV/c momentum are shown in  Fig. \ref{Fig:tracks-hadcal} (filled squares). They are compared to the DC1, DC2 analysis averaged over polar angles $0.03\le \theta\le0.24$\,rad (empty circles). The averaging produces a maximum value of $A_y$ that is slightly smaller than the corresponding maximum in Fig. \ref{Fig:p+CH2_3-42}.  

\begin{figure}[h]
\includegraphics [width=0.49\textwidth] {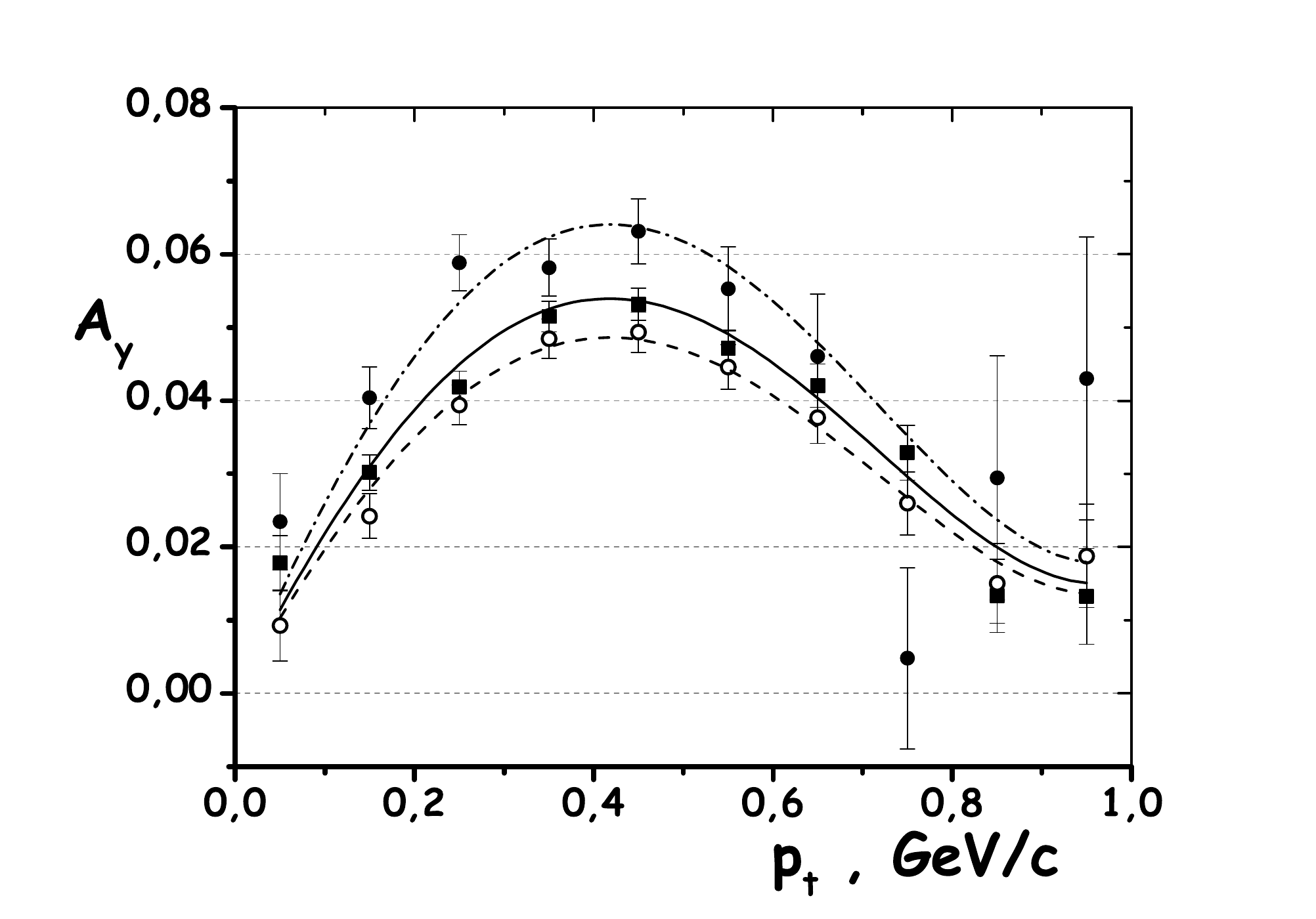}
\includegraphics [width=0.49\textwidth] {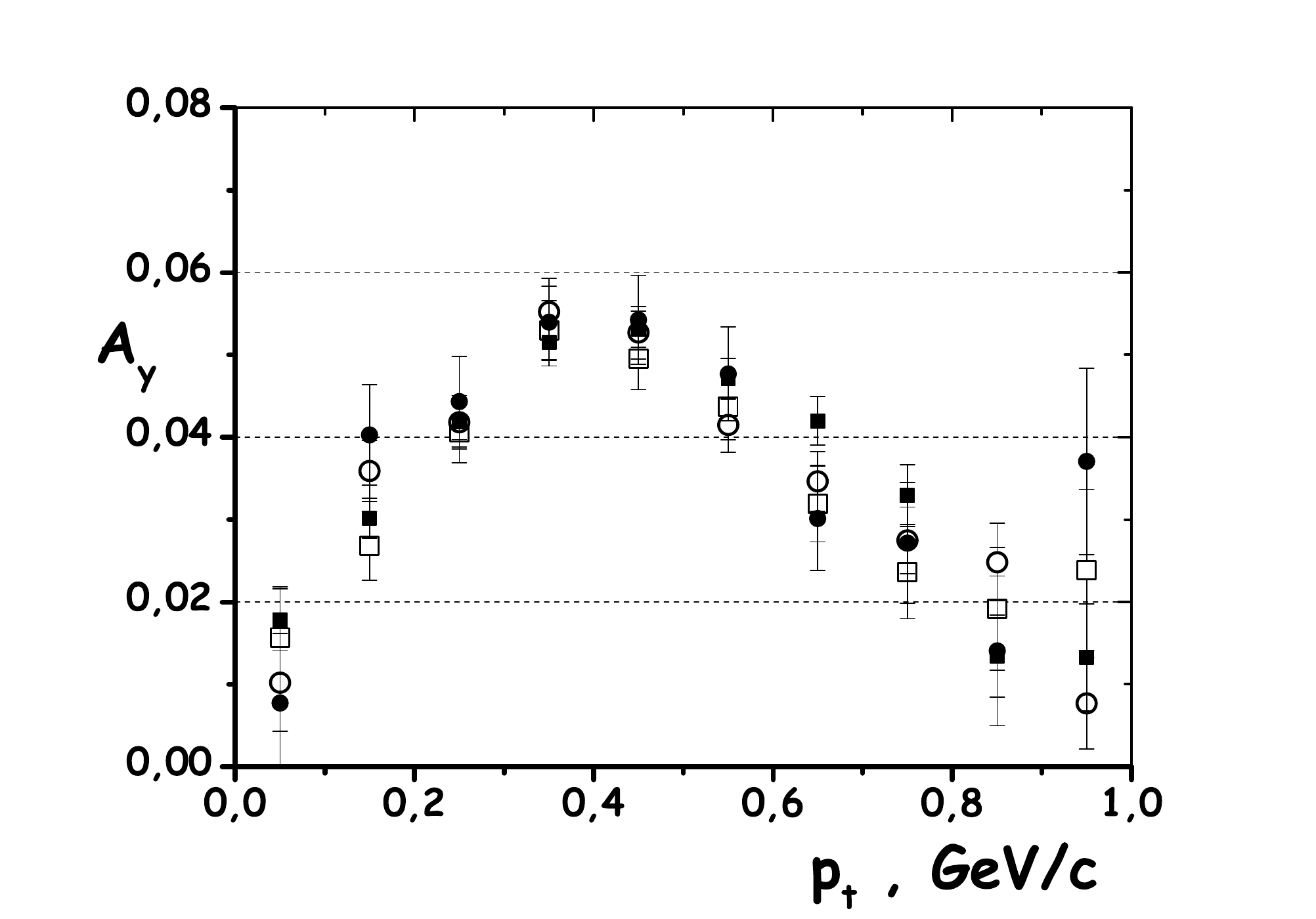}
\caption{$A_y$ as a function of $p_t$: (top) for $n+CH_2$ scattering at neutron different momenta: 3 GeV/c (solid circles),  3.75 GeV/c (solid squares), 4.2 GeV/c (open circles)- the lines represent eye guides through the data points; (bottom) for 3.75 GeV/c incident neutrons on different polarimeter targets: on $CH_2$  (solid squares), $CH$ (open circles), $C$ (open squares), and $Cu$ (solid circles).
}
 \label{Fig:n+CH2_3_42} 
\end{figure}

$A_y$ for the $n+CH_2$ charge exchange reaction (one final-state, charged particle is detected)
is shown in Fig. \ref{Fig:n+CH2_3_42} (top) at different momenta and in Fig. \ref{Fig:n+CH2_3_42} (bottom) for different polarimeter targets.
$A_y$ appears to decrease slightly with increasing neutron momentum, but the effect is much weaker than in $p+CH_2$ scattering. The dependence of $A_y$ on target material is very weak and there is no significant difference between data on $C$, $CH$, $CH_2$ and $Cu$.
%%: CH2 (black squares), CH (red circles) and C (green lozenges) as function of $p_t$.  One can see that  the analyzing powers decrease slowly with energy, Fig.  \ref{Fig:n+CH2_3_42}a. Increasing the hydrogen content of the target slightly increases the analyzing power, Fig.  \ref{Fig:n+CH2_3_42}b.

Charge exchange can also be studied for the process $p+CH_2\to n+X$, where neutral particles are detected with the condition that no charged track is recorded in DC1, DC2, but there is a suitably high energy deposit in the hadron calorimeter, from the interaction of a forward neutron.
 %%In this case one can not appreciate a large difference of analyzing powers with the incident proton momentum, Fig. \ref{Fig:p+CH2_n+X}.
The values of $A_y$ are similar to those obtained with a neutron beam and, contrary to the $p+CH_2\to p+X$ scattering case, there is little appreciable dependence of $A_y$ upon the the incident momentum. 

%In Fig. \ref{Fig:p+CH2_n+X} the product $A_yp_{lab}$ is shown as a function of $p_t$ for different beam momenta, in order to highlight the fact that the maximum analyzing power scales with the lab momentum of the incident beam, in the considered range.
 %\begin{figure}[h]
%\includegraphics
%[width=0.49\textwidth] {p+CH2_n+X.JPG}
%[width=0.49\textwidth] {Ay_n_x_mom.eps} 
%\caption{Analyzing power  times the proton lab momentum as a function of $p_t$ for $p+CH_2\to n+X$ scattering at 3 GeV/c (open circles), 3.75 GeV/c (solid squares),  and 4.2 GeV/c (crosses) incident momentum. The quantity $A_yP_{lab}$ shows the same maximum when varying the incident momentum.}
 %\label{Fig:p+CH2_n+X} 
%\end{figure}

\begin{figure}[h]
\includegraphics[width=0.49\textwidth] {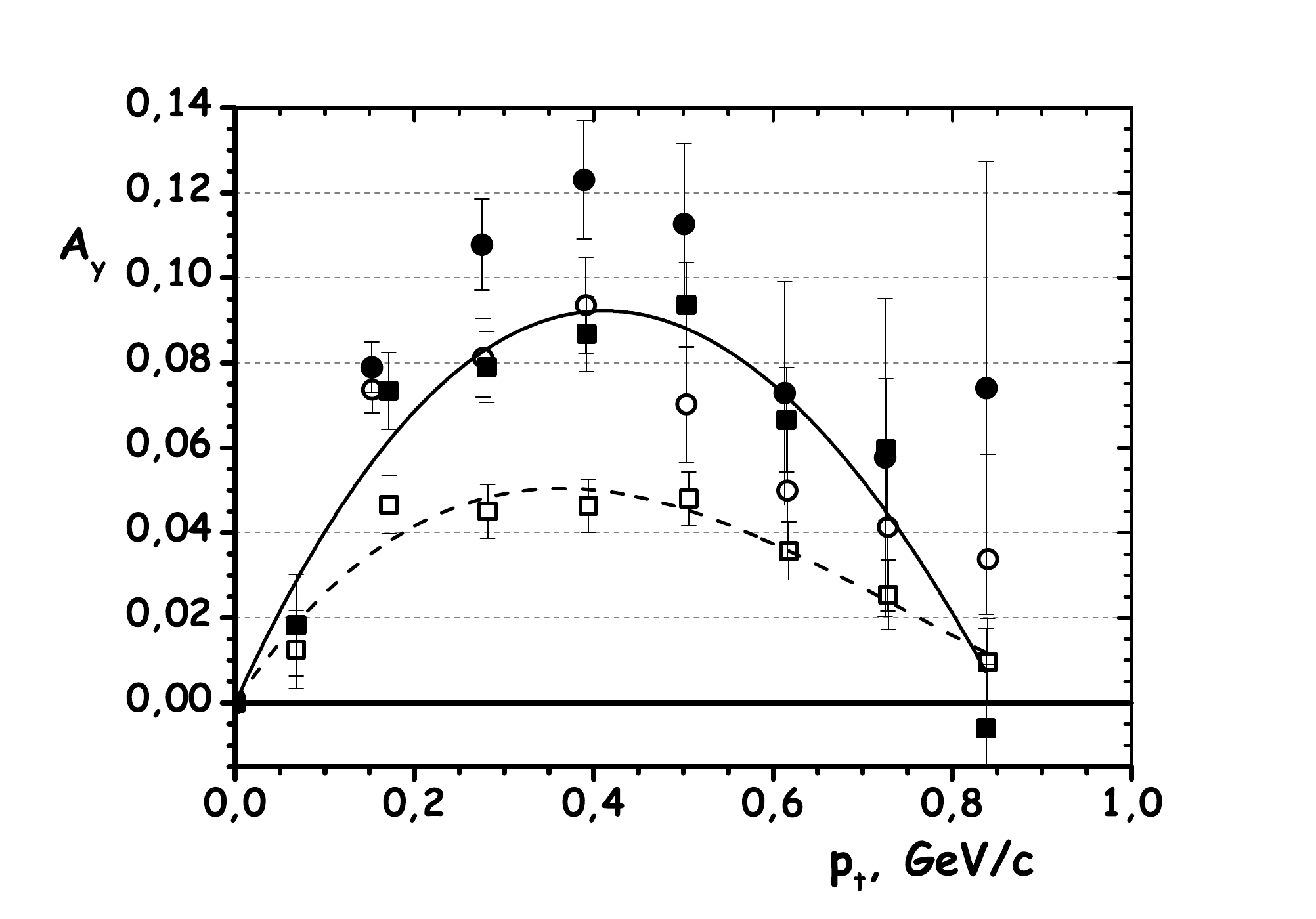} 
\caption{$p_t$ dependence of $A_y$ for quasi-elastic $p+Cu\to p+X$ as solid squares(open squares)  and charge-exchange $n+Cu\to p+X$ as open squares(open circles), for incident $p$ and $n$ momentum of 3.75 GeV/c before (after) selection of events with large energy deposit ($>6000$ [channels]) in the calorimeter. The lines represent eye guides through the data points for $n+Cu \to p+X$ before selection (dashed line) and after selection(solid line). }
 \label{Fig:p_n_Cu} 
\end{figure}

 For the first time, data were taken with 3.75 GeV/c polarized protons and neutrons incident on a $Cu$ target, detecting a forward-emitted charged particle. Fig. \ref{Fig:p_n_Cu} compares $A_y$ for charge-exchange $n+Cu\to p+X$ with the one for quasi-elastic $p+Cu\to p+X$. Where no threshold has been placed on the energy deposit from the hadron calorimeter, $A_y$ for $p+Cu$ is roughly double that for $n+Cu$. However, after selection of events with hadron calorimeter energy deposit greater than 6000 [channels], $A_y$ for $n+Cu$ increases by a factor $\sim 2$, while the increase for $p+Cu$ is $\sim$~1.3. 
%%The analyzing powers on a Cu target for neutron induced charge-exchange reaction is shown at 3.75 GeV/c in Fig.  \ref{Fig:p_n_Cu} (red solid squares) and it is compared with the (quasi) elastic reaction induced by a proton (blue solid squares). Proton (quasi)elastic scattering show twice as large maximum analyzing power. Fig. \ref{Fig:p_n_Cu} shows also that $A_y$  can be dramatically increased by selecting a large energy deposit in the hadron calorimeter. This effect is much larger for neutron charge exchange reaction, that shows 100\% increase for $nCu$ and 30\% increase for the $pCu$ reaction, both with  charged particles detected in the final state.
%%%%%%%%%%%%%%%%%%%%%%%%%%%
%% Text and caption were completely at odds with each other. I have assumed that the text was correct...jrma
%%%%%%%%%%%%%%%%%%%%%%%%%% 
%\end{enumerate}

%\end{enumerate}

%%%%%%%%%%%%%%%%%%
\section{Conclusions}
%%%%%%%%%%%%%%%%%%%%
Analyzing powers for polarized  protons and neutrons scattering on $C$, $CH$, $CH_2$ and $Cu$ targets were measured at nucleon momenta from 3.0 to 4.2 GeV/c by the ALPOM2 polarimeter at JINR Dubna. The unique polarized deuteron beam from the Nuclotron accelerator, which has a maximum momentum of 13 GeV/c, has been used to produce both polarized protons and (for the first time) polarized neutrons.

Analyzing powers have been obtained for elastic-like proton scattering and charge-exchange neutron scattering, both of which entail detection of a single, forward charged particle. Selection of high energy particles, using energy deposit in the calorimeter, is found to boost the analyzing power by a factor $\sim 2$ in the neutron case and $\sim 1.3$ in the proton case.
Analyzing powers have also been obtained for charge-exchange proton scattering, yielding values which are very similar to the neutron charge-exchange case.
The measured analyzer materials include $C$, $CH$, $CH_2$,and $Cu$, and show that $A_y$ in the multi-GeV/c domain is essentially the same for light and heavy nuclei. A heavy nucleus analyzer has the advantage of being much more compact for a given effective thickness of material.

Two polarimeter features, namely: the implementation of a calorimeter for the selection of high-energy, final-state nucleons,
and the replacement of a hydrogen-rich light target by heavier nuclei, open the way to simpler and more effective polarization measurements  for proton and neutrons in the GeV region.
Future experiments at JLab, requiring recoil-nucleon polarimetry, have already integrated these concepts, as in the case of experiments E12-07-109 \cite{PR12-07-109} and E12-17-004 \cite{PR12-17-004}.

These polarimetry concepts have also a range of applications at JINR. The inverse reaction $p+Cu$ (or $W$)$\to n+X$, with neutron detection in the forward direction by a hadron calorimeter, can be used for measurement of the proton polarization at the NICA collider. Spin effects in hadronic and heavy-ion collisions may be studied at NICA, using non-polarized, longitudinally and transversally polarized beams, constituting a consistent spin physics program \cite{Savin:2016arw}.

%with the ALPOM2 setup at the Nuclotron accelerator. The data with a  polarized neutron beam are obtained for the first time, thanks to the unique polarized deuteron beam that is presently available with momentum up to 13 GeV/c. The measurement of the angular dependence of $A_y$ for the neutrons  in the GeV region is essential to the continuation of neutron form factor measurements to the highest possible transferred momentum at JLab. 

%We showed that improving the selection of single tracks by energy deposit in a hadron calorimeter essentially increases the analyzing powers, for charge exchange as well as for elastic reactions. We also found that the reaction on a heavy target as Cu has large analyzing powers. This opens the way to a new concept of polarimetry in the GeV region, requiring much thinner and handball targets than the hydrogen-rich material  that were privileged until now. These are the main results of this work.

%\appendix
%\include{app-a}
\section{Appendix: Calculation of the analyzing powers}

The calculation of analyzing power is based on the analysis of the two $(\varphi,p_t)$-plots
corresponding to the two different polarization modes. We denote the bin values as $N_{ij}(1),N_{ij}(2)$,
where the first index, $'i'$, is related to $\varphi$ and  the second one, $'j'$, is related to $p_t$.

The values of the vector analyzing power $b_j$ are obtained from fit that minimizes the following quantity:
\begin{equation}\label{anpwv}
    \sum_{i,j} \left(\frac{R_{ij}-b_jf(\varphi)_i}{\Delta R_{ij}}\right)^2,
\end{equation}
with
\begin{eqnarray}
  \label{Rij}
  R_{ij} &=& \frac{N_{ij}(2)-C\cdot N_{ij}(1)}
   {C\cdot N_{ij}(1)P(2)+N_{ij}(2)P(1)},\\
  \Delta R_{ij} &=& \sqrt{\frac{N_{ij}(1)N_{ij}(2)}{[N_{ij}(1)+N_{ij}(2)]^3}}\cdot\frac{2}{|P(2)+P(1)|},
\end{eqnarray}
where
$P(k)$ refers to the beam polarization, $k=1,2$ relates to the polarization mode, and 
$$C=\frac{I[P(2)]}{I[P(1)]}$$
 is the ratio of intensities (I) in the different polarization modes. 

The fit is not equivalent to a number of $j$ independent fits: 
\begin{equation}\label{anpwvs}
    \sum_{i} \left(\frac{R_{ij}-b_jf(\varphi)_i}{\Delta R_{ij}}\right)^2 = min, 
\end{equation}
because the parameter $C$ is common for all $j$ and can be taken as a free parameter.
Such a fit provides reliable results due to the small correlation existing  between $C$ and $b_j$ in Eqs.~\ref{anpwv} and \ref{Rij} ($<0.5$\%).

This program is included into the latest FUMILIM package, Ref.~\cite{Sitnik:2016xxx}.

%
% BibTeX users please use
 %\bibliographystyle{unsrt}
 %\bibliography{Alpom}

%

\end{document}